\documentclass[traditabstract]{aa}

\usepackage{amsmath,amssymb}
\usepackage{subfigure}
\usepackage{graphicx}
\usepackage{color}
\usepackage{hyperref}
\usepackage{natbib} 
\bibpunct{(}{)}{;}{a}{}{,}

\newcommand{\Msun}{\rm M_{\odot}}

\newcommand{\nosfg}{g}
\newcommand{\sfg}{{g^{\star}}}

\newcommand{\mean}{\overline}

\begin{document}

\title{Galaxy stellar mass assembly: the difficulty matching
observations and semi-analytical predictions} 

\authorrunning{M. Cousin et al}

\titlerunning{Galaxy stellar mass assembly}

\author{M. Cousin \inst{1} \and G. Lagache \inst{1,4} \and M. Bethermin \inst{3} \and J. Blaizot \inst{2} \and B. Guiderdoni \inst{2}}

\institute{Institut d'Astrophysique Spatiale (IAS), B\^atiment 121, F- 91405 Orsay (France); Universit\'e Paris-Sud 11 and CNRS (UMR 8617) e-mail : morgane.cousin@ias.u-psud.fr \and Universit\'ee Lyon 1, Observatoire de Lyon, 9 avenue Charles Andr\'ee, Saint-Genis Laval, F-69230, France CNRS (UMR 5574), Centre de Recherche Astrophysique de Lyon, Ecole Normale Superieure de Lyon, Lyon, F-69007, France \and European Southern Observatory, Karl-Schwarzschild-Str. 2, 85748 Garching, Germany \and Aix Marseille Universit\'e, CNRS, LAM (Laboratoire d'Astrophysique de Marseille) UMR 7326, 13388, Marseille, France}

\date{Received ? / Accepted ?}

\abstract{Semi-analytical models (SAMs) are currently the best way to
understand the formation of galaxies within the cosmic dark-matter
structures. They are able to give a statistical view of the variety
of the evolutionary histories of galaxies in terms of star formation
and stellar mass assembly. 
While they reproduce the local stellar mass functions,
correlation functions and luminosity functions fairly well, they fail to match
observations at high redshift ($z\ge 3$) in most cases, particularly
in the low-mass range. The inconsistency between models and
observations indicates that the history of gas accretion in galaxies,
within their host dark-matter halo, and the transformation of gas
into stars, are not followed well. We briefly present a new version of the \textit{GalICS}
semi-analytical model. With this new model, we explore the impact of
classical mechanisms, such as supernova feedback or photoionization,
on the evolution of the stellar mass assembly and the star formation rate. Even
with strong efficiency, these two processes cannot explain the
observed stellar mass function and star formation rate distribution or
the stellar mass versus dark matter halo mass relation. We
thus introduce an ad hoc modification of the standard paradigm, based
on the presence of a no-star-forming gas component, and a
concentration of the star-forming gas in galaxy
discs. The main idea behind the existence of the
no-star-forming gas reservoir is that only a fraction of the
total gas mass in a galaxy is available to form stars. The reservoir
generates a delay between the accretion of the gas and the star
formation process. This new model is in much better agreement with the
observations of the stellar mass function in the low-mass range than
the previous models and agrees quite well with a large set of
observations, including the redshift evolution of the specific star
formation rate. However, it predicts a large amount of
no-star-forming baryonic gas, potentially 
larger than observed, even if its nature has still to be
examined in the context of the missing baryon problem. Outputs from all models are available online.}


\maketitle

\section{Introduction}
Cosmological models based on the $\Lambda$-CDM paradigm have proved
remarkably successful at explaining the origin and evolution of
structures in the Universe. Since the pioneer work of
\cite{Blumenthal_1984}, this model has become a powerful tool for describing
the evolution of primordial density fluctuations leading to the large
scale structures \citep[e.g.][]{Peacock_2001, Spergel_2003,
Planck_Collaboration_2013_Cosmological_parameters}. Galaxy clustering
or weak gravitational lensing are modelled very well in this
framework \citep[e.g.][]{Fu_2008}.\\

The description of smaller scales (galaxies) is more problematic.
Even if we put aside the problem of the angular momentum transfer
between the disc and the dark-matter host halo, there are still some
challenges on sub-galaxy scales. Twenty years ago,
\cite{Kauffmann_1993} pointed out the so-called sub-structure problem.
Indeed the large amount of power on small scales in the $\Lambda$-Cold-Dark-Matter(CDM)
paradigm generates an over-estimate of the number of small objects
(with properties close to dwarf galaxies). The over-density of
substructures is clearly seen in N-body simulations at low redshift
($z\simeq 0$). Dark matter haloes with mass comparable to that of our
Galaxy ($M_h \simeq 10^{12} \Msun$) contain more than one hundred substructures enclosed in their virial
radius. In contrast, the
observations of the Local Group count fifty satellite
galaxies at most.\\ 

This effect is even more problematic at high redshift ($z >
1$). Indeed, coupled with the poor understanding of the star
formation process in these small haloes, the standard scenario
produces a large excess of stellar mass in low-mass structures
\citep{Guo_2011}. To limit the number of dwarf galaxies, galaxy
formation models, such as semi-analytical model (SAM) or cosmological
hydrodynamic simulations, invoke gas photoionization and strong
supernova feedback \citep{Efstathiou_1992, Shapiro_1994, Babul_1992,
Quinn_1996, Thoul_1996, Bullock_2000, Gnedin_2000, Benson_2002,
Somerville_2002, Croton_2006, Hoeft_2006, Okamoto_2008,
Somerville_2008, Somerville_2012}. Originally proposed by
\cite{Doroshkevich_1967}, photoionization has been developed in
the $CDM$ paradigm by \cite{Couchman_1986}, \cite{Ikeuchi_1986}, and
\cite{Rees_1986}. The idea is quite simple: the ultraviolet (UV)
background generated by the quasars and first generations of stars
heats the gas. In the small structures, the temperature reached by
the gas is then too high, preventing it from collapsing into dark matter
haloes. The accretion of the gas on the galaxies, hence the star
formation, is thus reduced.\\

Many semi-analytical models strive to reproduce the luminous
properties of galaxy samples, such as luminosity functions or galaxy
number counts \citep{Cole_2000, Croton_2006, Hatton_2003,
Monaco_2007, Somerville_2008}. This approach has to be linked to the
nature of the observational constraints. Indeed, ten years ago,
broad-band luminosity measurements were the main constraints. In
general, local luminosity functions in the optical domain were well
reproduced by standard SAMs \citep{Cole_2000, Hatton_2003,
Croton_2006, Baugh_2006, Guo_2011}. But first analysis including the
dust reprocessing showed a deep misunderstanding of the star
formation processes \citep{Granato_2000}. 
Study of the cosmic infrared background, added to UV and optical
measurements, indicates a peak of star formation activity for $1<z<4$
\citep{Lilly_1996, Madau_1996, Gispert_2000, CharyElbaz_2001,
LeFloch_2005, Hopkins_2006, Dunne_2009, Rodighiero_2010a,
Gruppioni_2010}. The star formation rate distribution (or IR
luminosity function IR-LF) of galaxies at these epochs is currently
not reproduced well by the physical models \citep{Bell_2007,
LeFloch_2009, Rodighiero_2010a, Magnelli_2011}. Also discrepancies
between models and observations are large for the galaxy number
counts at long wavelengths ($\lambda > 100~\mu m$) or the redshift
distributions of star-forming galaxies \citep{Hatton_2003,
Baugh_2006, Somerville_2012}.
Prescriptions were developed to try to reduce the discrepancy, such
as the modification of the initial mass function (IMF) in starbursts
\citep{Guiderdoni_1997, Baugh_2006}. Even if thus a modification
improves the galaxy number counts in the far-infrared wavelengths
there is no observational evidence of such an IMF variation. \\

Today, with the new observational constraints, such as those derived
from galaxy-galaxy lensing \citep{McKay_2001,
Hoekstra_2004, Mandelbaum_2006a, Mandelbaum_2006b, Leauthaud_2010},
we have access to more fundamental galaxy properties: stellar mass
$M_{\star}$, star formation rates ($SFR$), and to the links between them
\citep{Brinchmann_2004, Noeske_2007, Dunne_2009, Elbaz_2011,
Karim_2011}. With the development of new techniques, such as the
abundance matching, the relation between stellar mass, galaxy mass,
dark matter halo mass, or even between $SFR$ and $M_{h}$ can be
explored \citep{Conroy_2009, Bethermin_2012a, Behroozi_2013b}.
Consequently, SAMs added some other relations to the analysis of the luminous properties
of galaxies, such as the specific star formation
rate ($sSFR = SFR/M_{\star}$) and its redshift evolution, or the
stellar mass ($M_{\star}$) versus dark matter halo mass ($M_{h}$)
relation (SHMR) \citep{Guo_2011, Leauthaud_2012}. The work presented
here continues this effort. \\

In this paper, we used a new semi-analytical model (detailed in
Cousin et al. 2014) built on recent theoretical prescriptions and
hydrodynamic simulation results \citep{Bertone_2005, Keres_2005,
Bournaud_2007, Genzel_2008, Dekel_2009a, Dekel_2009b, Khochfar_2009,
Voort_2010, Faucher-Giguere_2011, Lu_2011, Capelo_2012}, and we
compare it to an up-to-date set of observations. Our goal is to better understand the model parameters (physical
recipes) that have to be strongly modified to obtain good agreement
between models and observations. \\

The paper is organized as follows. In Sect.~\ref{model}, we
describe the main features of our SAM. In
Sect.~\ref{classical_results}, we explore the impact of classical
photoionization and supernova (SN)-feedback recipes on fundamental
galaxy properties: stellar mass function (SMF), $M_h$ versus
$M_{\star}$ relation (SHMR), and specific star formation rate (sSFR).
We add to these properties the SFR distribution (or
the IR-LF) and its redshift evolution. We
show that the basic models fail to reproduce these kinds of
measurements and propose the existence of a no-star-forming
gas reservoir in galaxy discs to reconcile the models with the
observations (Sect. \ref{ad-hoc}). We present a detailed comparison
between models and observations in Sect.~\ref{discussion}. We
conclude in Sect.~\ref{conclusion}. Throughout the paper we use \cite{Chabrier_2003}
initial mass function (IMF).

\section{Brief description of the model}
\label{model}
The SAM briefly presented here is a revised version of the GalICS
model \citep{Hatton_2003}. We did a detailed analysis of the
dark-matter merger tree properties and have revised the description
of baryonic physics using the most recent prescriptions extracted
from analytical works and hydrodynamic simulations. A
complete description is provided in a companion paper (Cousin et al., Towards a new modelling of gas flows in a semi-analytical model of galaxy formation and evolution).

\begin{table*}[t]
  \begin{center}
    \footnotesize{
      \begin{tabular}{lll}
        \hline
        Model & Definitions / Comments & Colour plots \\
        \hline
        $m_0$ & \cite{Okamoto_2008}, without (sn/agn)-feedback & red\\
        $m_1$ & \cite{Okamoto_2008} photoionization and our
(sn/agn)-feedback processes (\textit{reference}) & orange \\
        $m_2$ & \cite{Gnedin_2000} photoionization and our
(sn/agn)-feedback processes & green \\
        $m_3$ & \cite{Gnedin_2000} photoionization and
\cite{Somerville_2008} SN-feedback, without AGN-feedback & cyan \\
        $m_4$ & \textit{reference} +
\textit{\textit{no-star-forming}} gas disc component (Sect.
\ref{ad-hoc-2}) & purple \\
        \hline
      \end{tabular}}
  \end{center}  
  \caption{\footnotesize{List of SAMs compared in this paper.}}
  \label{model_description}
\end{table*}

\subsection{Dark matter}
\label{dm}

Like its predecessor, our model is based on a hybrid approach. We use
dark-matter merger trees extracted from a pure N-body simulation.
This simulation, with WMAP-3yr cosmology ($\Omega_m = 0.24$,
$\Omega_{\Lambda} = 0.76$, $f_b = 0.16$, $h = 0.73$), describes a
volume of $(100h^{-1})^3 \simeq 150~Mpc^3$. In this volume, $1024^3$
particles evolve with an elementary mass of $m_p = 8.536~10^7~\Msun$.
We use the \verb?HaloMaker? code described in \cite{Tweed_2009} to
identify the haloes and their sub-structures, and build merger trees.
We only consider dark-matter structures containing at least 20
dark-matter particles. This limit gives a minimal dark matter halo
mass $M_h^{min} = 1.707\times10^9~\Msun$.\\

In addition to the merger-tree building, we have added a post-treatment to
the dark-matter haloes. Based on the time-integrated halo mass and on
the energy and halo spin parameter evolution, we selected the
\textit{healthy} population of haloes, i.e., haloes with a negative
total gravitational energy and a smooth evolution of the spin
parameter. The tree branches that do not satisfy the conditions
are considered as smooth accretion ($\simeq 1-5\%$ of the
total mass identified in haloes at a given time). There are no
galaxies in these kinds of tree branches.

\subsection{Adding baryons}
 \label{add_baryons}

\begin{table}[h]
  \begin{center}
    \footnotesize{
      \begin{tabular}{lll}
        \hline
        Symbol & Definition & Value \\
        \hline
        $\varepsilon_{\star}$  & Star formation efficiency [Eq.
\ref{star_formation_law}] & 0.02 \\
        $\varepsilon_{ej}$  & SN feedback efficiency [Eq.
\ref{sn_feedback}] & 0.3 \\
        $\tau_{merger}$  & Merger time scale [Eq. \ref{boost_factor}]
& 0.05 Gyr \\
        $\left<f_b\right>$  & Universal baryonic fraction [Eq.
\ref{ph-ion}] & 0.18 \\
        \hline
      \end{tabular}}
  \end{center}  
  \caption{\footnotesize{List of the main models parameters. The
values given here are identical or very similar to those commonly used
in the literature.}}
  \label{model_parameters}
\end{table}

In hybrid SAMs, the baryonic physics are added to the pre-evolved
dark-matter background. The baryonic mass is added progressively,
following the dark-matter smooth accretion: $\dot{M}_b =
f_b^{ph-ion}(M_h,z)\dot{M}_{dm}$, where $f_b^{ph-ion}(M_h,z)$ depends
on the photoionization model. In our case, we use the
\cite{Okamoto_2008} prescription in the reference model $m_1$ (see
Sect. \ref{ph-io_impact} for more information), and we use
\cite{Gnedin_2000} as a model variation.\\

In the current galaxy formation paradigm, the baryonic accretion
that leads to the galaxy formation can be separated into two different
phases \citep[e.g.][]{Keres_2005, Dekel_2009a, Dekel_2009b,
Khochfar_2009, Voort_2010, Faucher-Giguere_2011}. On the one hand,
we distinguish a cold mode where the gas is accreted through
the filamentary streams. The cold mode dominates the growth of galaxies at high redshifts, and the growth of lower mass
objects at any times. On the other hand, in more massive haloes
($M_{h} > 10^{12}~\Msun$) and at low $z$, the accretion is dominated
by a hot mode, where a large fraction of the gas is
shock-heated to temperatures close to the virial temperature. This
gas feeds a hot stable atmosphere ($T_g > 10^5~K$) around
the central host galaxy. To take into account this bimodal accretion,
we use \cite{Lu_2011} prescription (their Eqs. 24 and 25). The
accreted mass, divided into the two modes, is stored in two different
reservoirs $M_{cold}$ and $M_{hot}$. The two reservoirs feed the
galaxy with rates close to the free-fall rate for the cold mode and
follows a cooling process for the hot mode.

\subsection{Disc formation}
\label{disc_formation}
Accretion from cold streams and cooling flows feed the galaxy disc
in the centre of the dark-matter halo. We assume that this cold gas
initially forms a thin exponential disc. Gas acquires angular
momentum during the mass transfer \citep{Peebles_1969}. After its
formation, the disc is supported by its angular momentum. This
paradigm is based on the prescription given by \cite{Blumenthal_1986}
or \cite{Mo_1998}, and has been frequently used in SAMs, as in \cite{Cole_1991}, \cite{Cole_2000}, \cite{Hatton_2003}, or
\cite{Somerville_2008}.\\

Since more than one decade, observations \citep{Cowie_1995,
Van_den_Bergh_1996, Elmegreen_2005, Genzel_2008, Bournaud_2008} and
hydrodynamic simulations
\citep{Bournaud_2007,Ceverino_2010,Ceverino_2012} show the existence
of gas-rich turbulent discs at high $z$. These discs are unstable
and undergo gravitational fragmentation that forms giant
clumps. These clumps interact and migrate to the centre of the galaxy
where they form a pseudo-bulge component
\citep{Elmegreen_2009, Dekel_2009b}.
In our model, we use a new self consistent model of disc
instabilities. We assume that, in the disc, mass over-density and
low-velocity dispersion lead to the formation and migration of
giant clumps. A complete description of this process, which
is based on \cite{Dekel_2009b} and which has been adapted to our SAM
approach, is given in a companion paper (submitted to A\&A).
In brief, we compute the instantaneous unstable disc mass using the
Toomre criterium \citep{Toomre_1963, Toomre_1964}. This unstable mass
(mass in clumps) increases with time following the evolution of the
disc. When this mass becomes higher than a mass threshold
corresponding to a characteristic individual clump mass
\citep{Dekel_2009b}, we compute the transfer of the clump mass from
the disc to the pseudo-bulge component. This transfer is
modelled as a micro-merger event with the pre-existing bulge
component.

\subsection{Star formation}
\label{star_formation}

In each galaxy component, disc and/or bulge, the cold gas mass
$M_\sfg$ is converted into stars. In standard models, the totality of
the cold gas can be converted into stars. In Sect.~\ref{ad-hoc-2}, we present a strong modification of this prescription by
introducing a no-star-forming gas component
($M_{\nosfg}$). In anticipation to this change, we specify here that
obviously only the star-forming gas component ($M_\sfg$)
takes part in the star formation process. We use the following
standard definition of the SFR:
  \begin{equation}
    \dot{M}_{\star} = \varepsilon_{\star}\dfrac{M_\sfg}{t_{dyn}}
    \label{star_formation_law}
  \end{equation}
where $\varepsilon_{\star}~(=~0.02)$ is a free parameter adjusted to
follow the Schmidt-Kennicutt relation \citep{Kennicutt_1998}, and
$t_{dyn}$ is the dynamical time given by
\begin{equation}
t_{dyn} = \left\{
  \begin{array}{ll}
   MIN\left(2r_{1/2}\sigma_v^{-1}~,~2\pi r_{1/2} V_c^{-1}\right)  &
\mbox{: for disc} \\
    & \\
    2r_{1/2}\sigma_v^{-1} & \mbox{: for bulge}
  \end{array}\right.
\label{t_dyn}
\end{equation}
where $V_c$ is the circular velocity measured at the half radius
mass, and $\sigma_V$ is the mean velocity dispersion. For
completeness, we add that the star formation is computed only if the
projected star forming gas surface density $\Sigma_g$ is higher than
a given threshold $log_{10}(\Sigma_g^{min})~=~1~[\Msun\cdot pc^2]$.  

\subsection{Supernovae feedback}
\label{sn_feedback}

In a given stellar population, massive stars evolve quickly and end
their life as supernovae. This violent death injects gas and energy
into the interstellar medium. The gas is heated, and a fraction can be
ejected from the galaxy plane and feed the surrounding host-halo
phase. In addition, these ejecta are at the origin of the metal
enrichment of structures. Supernova feedback is therefore a crucial
ingredient. In the majority of SAM \citep[e.g.][]{Kauffmann_1993,
Cole_1994, Cole_2000, Silk_2003, Hatton_2003, Somerville_2008}, and
according to some observational studies \citep[e.g.][]{Martin_1999,
Heckman_2000, Veilleux_2005}, the SN-reheating or
SN-ejecta rate is linked to the SFR. As
proposed by \cite{Dekel_1986}, we computed the ejected mass rate due
to supernovae by using kinetic energy conservation. Another paradigm
based on momentum conservation could be used, but it has been shown
by \cite{Dutton_2009} that, in the low-mass regime, the energy-driven
feedback is more efficient and leads to better results.\\

The ejected mass rate $\dot{M}_{ej,SN}$ due to SN is linked to the
SFR $\dot{M}_{\star}$ by using the individual
supernova kinetic energy as:
\begin{equation}
    	\dot{M}_{ej,SN}V_{wind}^2 =
2\varepsilon_{ej}\eta_{sn}E_{sn}\dot{M}_{\star}
        \label{sn_feedback}
\end{equation}
where we use $\eta_{sn}=9.3\times 10^{-3}~\Msun^{-1}$ and an
efficiency $\varepsilon_{ej} =0.3$\footnote{The influence of the
efficiency value has been tested in the range
$\varepsilon_{ej}\in[0.05,10]$. Obviously a strong increase in the
SN-efficiency increases the amount of ejected gas. The star formation
activity is therefore reduced, but this effect affects only the
amplitude and not the shape of the stellar mass function. Moreover,
looking at the amplitude, its decrease is not enough to be in
agreement with the observations.}. The value used for this parameter
is similar to the one applied in standard SAMs
\citep[e.g.][]{Somerville_2008, Guo_2011}.\\

To break the degeneracy between the ejected mass and the velocity of
the wind, we must add a constraint on the wind velocity. We rely on
\cite{Bertone_2005} in which the wind velocity is linked to the
star formation rate \citep{Martin_1999}. It seems to be independent
of the galaxy morphology \citep{Heckman_2000, Frye_2002}. We
therefore use Eq. 9 in \cite{Bertone_2005} to model the wind
velocity.\\

On average, wind velocities obtained with this prescription are
higher than in other studies
\citep[e.g.][]{Somerville_2008,Dutton_2009}. Indeed it is common to
use galaxy escape velocity to describe the wind, which is, for the
ejection process, the minimum required value. Therefore, the ejected
mass is at its maximum (see \cite{Dutton_2009}, their discussion in
Sect.7.3). Consequently, our loading factor
($\dot{M}_{ej,SN}/\dot{M}_{\star}$) is smaller than in other models,
and therefore our mean ejected mass is also lower. The difference
between our reference model and standard supernova feedback is
discussed in Sect.~\ref{sn_feedback_impact}.

\subsection{The active galaxy nucleus}
\label{agn_component}

A supermassive black hole (SMBH) can evolve in the centre of
the bulge. We form the seed of the SMBH by converting a fraction of
the bulge mass (gas and stars) to the SMBH mass, when the bulge mass
becomes higher than a mass threshold $M_{bulge} \ge
10^3M_{bh}~\Msun$. This formation process is only turned on during a
merger event. The SMBH formed at this time has a mass equal to
$M_{bh} = 10^{3}~\Msun$. This mass is created by instantaneously converting a fraction of the gas and stars in their respective
ratio. 
After its formation, the SMBH evolves in the centre of the
bulge by accretion and clumps migration. The accretion process is
mainly driven by the \citep{Bondi_1952} accretion prescription, and we
add an episodic accretion linked to
clumps migration to this classical mechanism. This accretion is obviously limited by the maximum
value of the Eddington accretion rate.
As for supernovae, AGNs produce winds and contribute to the
hot-atmosphere heating. We convert a given fraction of the power
produced by mass accretion into kinetic and thermal power ($f_{Kin} =
10^{-3}$, \citep[e.g.][]{Proga_2000, Stoll_2009, Ostriker_2010}. To compute the AGN ejected mass rate, we
use the same kinetic conservation criterium as applied to
supernovae: $\dot{M}_{jet}V_{jet}^2 \propto \dot{M}_{agn}c^2$. Then
we compute the momentum transfer between the AGN jet and the gas to
estimate the mass that leaves the galaxy owing to AGN/gas coupling. We
assume that all the mass is ejected with a velocity equal to the
galaxy escape velocity
($\dot{M}_{ej,AGN}\propto\dot{M}_{jet}\frac{V_{jet}}{V_{esc}}$). As
explained in Sect. \ref{hot_phase}, the thermal power of the AGN is
used for the monitoring of the hot phase temperature.
 
\subsection{Hot-halo phase}
\label{hot_phase}

As mentioned previously, galaxies hosted by massive dark matter
haloes present a hot stable atmosphere generated by the hot
cosmological accretion and maintained by galaxy ejecta. We use a new
self consistent model for the hot halo phase evolution. We follow the
hot halo phase mean temperature ($\mean{T}$) by applying a
conservation criterion on the energy produced by hot accretion and/or
feedback winds coming from SN/AGN. For SN and AGN, a fixed fraction
($f_{Therm} = 5\%$) of the non-kinetic energy is converted into wind-thermal energy.\\ 

In parallel to the mean temperature monitoring, we compute the
evaporated mass\footnote{The hot atmosphere is considered in
hydrostatic equilibrium in the dark matter halo potential well. We
use a Maxwell-Boltzmann function to describe the velocity
distribution. At each time step, the mass corresponding to the hot gas
that has higher velocities than the escape velocity of the halo is
definitively removed.} and mass loss due to galactic winds. All these
calculations are done in a dominant dark-matter gravitational
potential and assume the hydrostatic equilibrium (HEC;
\citealt{Suto_1998, Makino_1998, Komatsu_2001, Capelo_2012}). This
new model gives a mean temperature ($\mean{T}$) close to the standard
virial temperature prescription for structure in the intermediate
dark matter halo mass range ($10^{10}-10^{12}~\Msun$). Low-mass
structures have higher temperatures ($\le 2T_{vir}$). This result is
linked to the gas ionization fraction that is assumed to be the same
for all structures. Massive structures that host an active galaxy
nucleus also have higher mean temperatures, but never higher than
three times the temperature derived from the standard virial
assumption.

\subsection{Cooling processes}

Cooling is computed using the classical model initially proposed by
\cite{White_1991}. The condensed mass enclosed in the cooling radius
$r_{cool}$ is estimated assuming
\begin{itemize}
  \item{an HEC gas profile $\rho_g(r)$ \citep{Suto_1998, Makino_1998,
Komatsu_2001, Capelo_2012},}
  \item{a mean constant temperature $\overline{T}$,}
  \item{a temperature and metal dependent cooling function
$\Lambda(T,Z_g)$ \citep{Sutherland_1993}.}
\end{itemize}
The cooling radius $r_{cool}$ is the unique solution for $t(r_{cool})
= t_{cool}$, where $t_{cool}$ is the effective cooling time computed
as the life time of the hot gas phase, and $t(r)$ is the cooling time
function of \cite{White_1991}
\begin{equation}
    t(r) = 0.64\dfrac{m_p k_b
\overline{T}}{\rho_g(r)\Lambda[\overline{T},Z_g]}.
    \label{cooling_time_function}
\end{equation}

\subsection{Mergers and bulge growth}
\label{merger}

At a given time step $t_n$, if two or more haloes have the same
descendant at $t^{n+1}$, these haloes and their host galaxies have
merged during this time lapse. We assume that the merger occurs
at $t_{merge} = 0.5\times(t^{n+1}+t_n)$\footnote{However, even if we
accurately follow the sub-haloes and if the merging time laps is
reduced, the instantaneous merging of galaxies is always a strong
assumption and could be explored in detail in a future work.}.
Between $t_n$ and $t_{merge}$, the progenitors evolve in their
host dark-matter halo and between $t_{merge}$ and  $t^{n+1}$ the remnant galaxy evolves in the descendent dark-matter halo. Even
if more than two progenitors are identified, mergers are
computed using dark-matter (and associated galaxy) pairs starting
from the lower sub-halo mass to the higher main halo mass. The
post-merger galaxy morphology depends on the mass (galaxy +
dark-matter halo) ratio of the two progenitors. We define
\begin{equation}
    \eta_{merger} =
\dfrac{MIN(M_{1/2,1}~;~M_{1/2,2})}{MAX(M_{1/2,1}~;~M_{1/2,2})}
    \label{mu_merger}
\end{equation}
where $M_{1/2,i} = M_{gal,i}(r<r_{1/2}) + 2M_{dm,i}(r<r_{1/2})$ is,
for system $i$, the sum of the galaxy and the dark matter halo
mass enclosed in the galaxy half-mass radius ($r_{1/2}$).\\

For $\eta_{merge} < 0.25$, we consider that it is a minor merger. In this
case, the disc and the pre-existing bulge component are kept, and gas
and star contents are just added. The remnant disc size is set to
the larger disc progenitor size. We apply the same rule to the bulge
component. The velocities (dispersion and circular) of the bulge and
the disc are recomputed with the properties of the remanent
dark-matter halo. 
In the case of a major merger, $\eta_{merge} > 0.25$, progenitor discs
are destroyed, and the remanent galaxy is only made of a bulge. The
half mass radius of this spheroid is computed using the energy
conservation and the virial theorem (as in \citealt{Hatton_2003}).
Like a pseudo-bulge component formed by \textit{giant}
clumps migration (Sect.~\ref{disc_formation}), bulges are described
by a \cite{Hernquist_1990} model.
After a major merger event, all new accreted material generates a new
disc component. The mass is only transferred to the bulge by disc
instabilities (Sect. \ref{disc_formation}).

Mergers are violent events. On a short time scale, the galaxy
properties are strongly modified, and secular evolution laws
(efficiencies) are no longer valid. To take the
modifications induced by a merger into account, we use a boost factor,
$MAX\left[1,~\varepsilon_{boost}(\Delta t)\right]$, that increases
the efficiencies of star formation in each component of the galaxy
(disc and bulge), and of SMBH accretion. The boost factor is defined as: 
\begin{equation}
    \varepsilon_{boost}(\tau) =
100\eta_{merger}\eta_{gas}exp\left(-\dfrac{\tau}{\tau_{merger}}\right).
    \label{boost_factor}
\end{equation}
where $\eta_{gas}$ is the gas fraction in the post-merger structure,
$\tau$ is the time elapsed since the last merger event,
$\tau_{merger}$=0.05~Gyr is a characteristic merger-time scale and
$\eta_{merger}$ is given in Eq.~\ref{mu_merger}. The formulation is
used to simulate a time-dependent gas compression (decreasing with
time) and takes the gas content of the two progenitors into account. 
The more gas they contain, the more the gas compression is high.  

\subsection{The adaptive time-step scheme}

In a galaxy, various processes act at the same time on various time
scales. For example, it is not efficient to compute the evolution of
the cold filamentary phase, which evolves on a typical dark-matter
dynamical time ($10^6$ yr), with a time step following the ejection
rate of the galaxy ($10^4$ yr). Using the same time step for all
components generates numerical errors on the slowly evolving
component and degrades the precision. Each component of the baryonic
halo or of the galaxy (disc and/or bulge) must evolve with a time
step that is as close as possible to its dynamical time. Accordingly,
we have developed an adaptive time-step scheme. Each halo or galaxy
component has a separated evolution scheme, and interacts with others
only if the mass transfers significantly affect its evolution. We
consider that the mass reservoir is modified if the variation is
over $10\%$.

\section{Star formation in the low-mass structures in standard models}
\label{classical_results}

\begin{figure}[ht]
  \begin{center}
    \includegraphics[scale =
0.92]{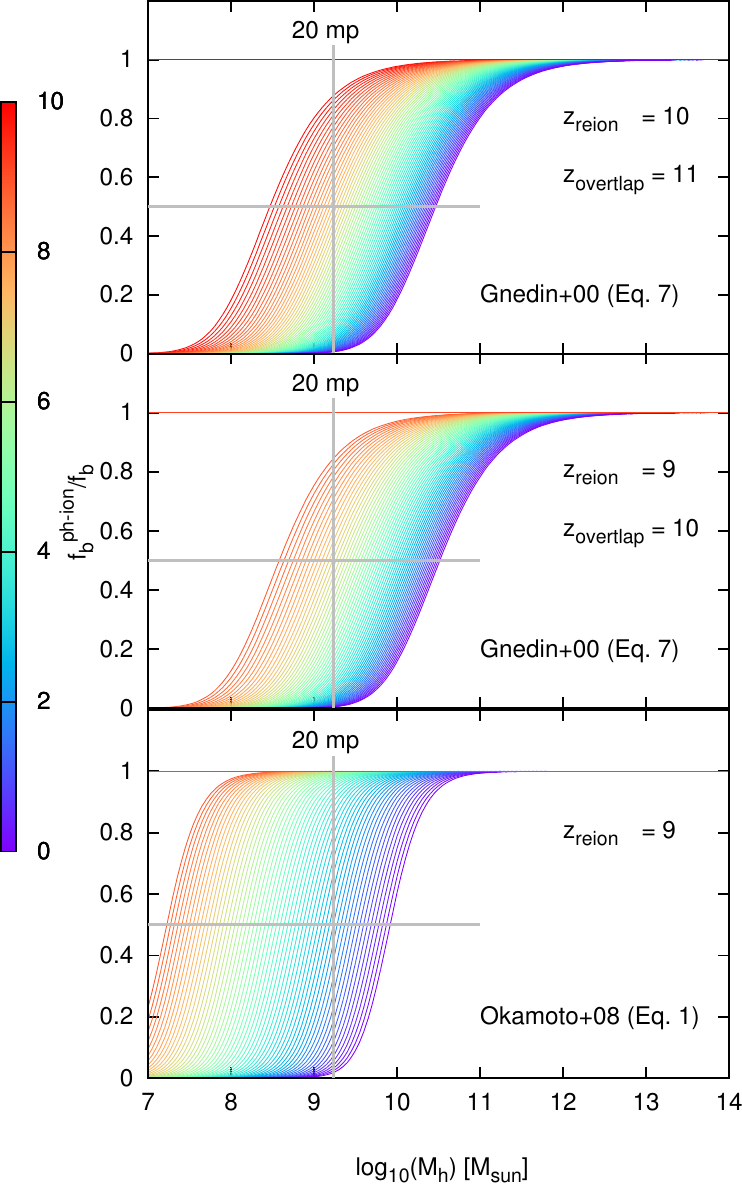}
  \caption{\tiny{Normalized baryonic fraction $f_b^{ph-ion} /f_b$ as
a function of the dark matter halo mass. The colour code shows the
redshift evolution (from z = 10 in red, to z = 0 in purple). We
compare two different prescriptions for the evolution of the baryonic
fraction after the redshift of reionization. The two upper panels are
made using the \cite{Gnedin_2000} prescription with two different
reionization redshifts (zreion = 10 and 9). The bottom panel shows the
\cite{Okamoto_2008} prescription with zreion = 9. The grey vertical
line indicates our dark-matter mass resolution limits (20
dm-particles). In the \cite{Gnedin_2000} model (central panel) the strong
effects induced by photoionization ($f_b^{ph-ion} <
0.5\left<f_b\right>$) appear at $z \simeq 7-8$ and become stronger
with decreasing redshift. In the \cite{Okamoto_2008} prescription, the
impact of photoionization comes much later ($z < 1$) and cannot
affect the galaxy formation processes at high $z$.}}
  \label{photoionization_model}
  \end{center}
\end{figure}

We focus in this section on the main problem of star formation
activity in low-mass structures. We explore various prescriptions for
the photoionization or SN-feedback processes and compare the
results with some fundamental galaxy properties: stellar mass
function (SMF), $M_h$ versus $M_{\star}$ relation (SHMR), the sSFR
versus $M_{\star}$ relation, and SFR distribution (or
IR-LF). \\
Table~\ref{model_description} gives the description of the models,
from $m_0$ without feedback processes (in red) to $m_4$, where we put
a large fraction of the gas into a no-star-forming gas
component (see Sect.~\ref{ad-hoc-2}). 
In the low-mass range analysed here, AGN-feedback does not play an
important role, and therefore it is not further discussed.

\subsection{Impact of photoionization}
\label{ph-io_impact}

Gas heating generated by the first generation of stars and quasars
limits the baryonic gas accretion in the smaller structures
\citep{Kauffmann_1993}. The effective baryonic fraction
$f_b^{ph-ion}$ (Eq. \ref{ph-ion}) depends on both the redshift and
the dark matter halo mass. The most commonly used formulation is the one
proposed by \cite{Gnedin_2000} and \cite{Kravtsov_2004}: 
  \begin{equation}
   f_b^{ph-ion}(M_h,z) = \left<f_b\right>\left[1 +
(2^{\alpha/3}-1)\left(\dfrac{M_h}{M_c(z)}\right)^{-\alpha}\right]^{-3/\alpha}.
   \label{ph-ion}
  \end{equation}
In this definition, $\left<f_b\right>$ is the universal
baryonic fraction, $M_h$ the dark matter halo mass, and $M_c(z)$ the
filtering mass corresponding to the mass where the halo lost half of
its baryons. Finally, $\alpha$ is a free parameter that mainly
controls the slope of the transition.
\begin{itemize}
  \item{For $\alpha=1$,$f_b^{ph-ion}/\left<f_b\right>: 0\rightarrow
1$ for M$_h: 10^9 \rightarrow 10^{12}$ M$_{\odot}$,}
 \item{For $\alpha=2$, $f_b^{ph-ion}/\left<f_b\right>: 0\rightarrow
1$ for M$_h: 10^9 \rightarrow 10^{10}$ M$_{\odot}$.}
\end{itemize}
 The redshift evolution of the filtering mass $M_c$ and the value of
$\alpha$ are the crucial parameters governing the impact of
photoionization on small structures.\\

These parameters have been constrained using hydrodynamic simulations that
include UV photoionization performed by, for example,
\cite{Gnedin_2000}, \cite{Kravtsov_2004}, \cite{Hoeft_2006}, or
\cite{Okamoto_2008}. The analysis of these different simulations
gives different results, and therefore various parameter values and
filtering mass behaviours. While in \cite{Gnedin_2000} the slope
index is set to ($\alpha = 1$), \cite{Okamoto_2008} find a higher
value ($\alpha = 2$). We recall in Appendix~\ref{filtering _mass} the
mathematical expressions for the two filtering masses given in these
papers and used here.\\

In Fig~\ref{photoionization_model}, we show  the evolution of
$f_b^{ph-ion}/\left<f_b\right>$ for the two prescriptions. The two
top panels are dedicated to the \cite{Gnedin_2000} model. Their
filtering mass definition is the one most commonly used in the literature
\citep{Somerville_2002, Croton_2006, Somerville_2008, Guo_2011,
Somerville_2012}. In the upper panel, we apply $z_{reion} = 10$, as
in \cite{Somerville_2012}. In the central panel we apply $z_{reion} =
9$ to compare with \cite{Okamoto_2008}. The grey horizontal line
marks a decrease of 50\% comparing to the universal baryonic fraction. We
consider that the photoionization effect is important when
$f_b^{ph-ion} < 0.5 \left<f_b\right>$. The colour code indicates the
redshift evolution.\\

At our dark matter halo mass resolution, in the first case
(\citealt{Gnedin_2000}, $z_{reion} = 10$), photoionization starts to
play a role at $z \simeq 8$. At low $z$, the small structures are
strongly affected by the photoionization process. This trend is
still true when the reionization redshift is decreased (central
panel, $z_{reion} = 9$). The effect of photoionization is much less
important in the \cite{Okamoto_2008} prescription. Indeed, the
significant decrease in $f_b^{ph-ion}/\left<f_b\right>$ only appears at
the mass resolution for redshift $z<1$. In this case, the gas
heating due to the UV background cannot affect, at high redshift, the
baryonic assembly of small structures. This difference in behaviour
comes from the different redshift evolutions of the filtering mass
$M_c(z)$. In the two cases, the authors use hydrodynamic simulations
to constrain this evolution. As explained by \cite{Okamoto_2008},
even if the two studies use different assumptions (e.g. link between
gas density and gas temperature), it seems that the difference
between the two prescriptions is most likely due to insufficient
resolution of the \cite{Gnedin_2000} simulation.\\ 

We applyed the two photoionization prescriptions in models $m_1$ and
$m_2$. The first one, $m_1$, uses the \cite{Okamoto_2008} description. It
is our reference model. For comparison, we use \cite{Gnedin_2000}
prescription in $m_2$. As we can see in Figs.~\ref{mass_function} and
\ref{Mhalo_Mstars}, the two models mainly have an impact at low
redshift. The figures show that the \cite{Gnedin_2000} prescription
reduces the stellar mass formed in the small dark matter mass
regime more than does the \cite{Okamoto_2008} prescription. This is consistent
with the baryonic fraction behaviour described above. Gas accretion
is more reduced in the \cite{Gnedin_2000} model. At low halo mass
($M_{halo} < 5\times 10^{10}~\Msun$) and at low redshift ($z < 2$),
the mean stellar mass built through the \cite{Gnedin_2000}
photoionization model ($m_2$) may be ten times lower than the one built
with the \cite{Okamoto_2008} model ($m_1$).\\

Currently, the majority of SAMs use the \cite{Gnedin_2000}
photoionization parameterization. The parameter set ($\alpha$,
$M_c(z)$) used in this case leads to an accretion rate on the
galaxies that is reduced compared to what is obtained using
\cite{Okamoto_2008}. This result fully agrees with
\cite{Guo_2011}. However, regardless of the case, it is evident from
Fig.~\ref{mass_function} that photoionization is not enough to
reduce the low-mass end of the stellar mass function as required by
the observations. 

\subsection{Impact of SN feedback}
\label{sn_feedback_impact}

While the photoionization process reduces the gas feeding of the
galaxy, supernova feedback expels the gas already present in the
galaxy. Despite their different actions, both processes tend to
reduce the amount of gas available to form stars.\\
We compare two SN-feedback models:  
\begin{itemize}
  \item{our model based on kinetic, thermal energy conservation, hot
gas phase heating, and evaporation;}
  \item{the \cite{Somerville_2008} model based on their Eqs.~12 and
13 (reheated rate and escape fraction). In this case, the hot gas-phase temperature is not monitored as it is in our model, but is set
to the dark matter virial temperature.
  }
\end{itemize}
As listed in Table \ref{model_description}, $m_2$ and $m_3$ used
the same photoionization prescription (\cite{Gnedin_2000}). 
They differ only in their SN-feedback model. A simple comparison
between the stellar mass functions (Fig. \ref{mass_function}) given
by $m_2$ and model $m_3$ in which we have implemented the
\cite{Somerville_2008} prescription indicates that the SN-feedback
mechanism is more efficient in their model. This difference is even
more visible on the SHMR (Fig. \ref{Mhalo_Mstars}) which indicates
that the stellar mass produced in low-mass haloes ($M_h <
10^{11}~\Msun$) is, on average, higher in our model by a factor close
to 3. This difference decreases when $z$ decreases and $M_h$
increases.\\

As explained previously (Sect.~\ref{sn_feedback}), the higher mean
wind velocity computed in our model following \cite{Bertone_2005}
leads to a lower ejected mass for a given kinetic energy. A more
detailed comparison of the two models proves that the reheating rate
computed with the \cite{Somerville_2008} SN-feedback model ($m_3$) is,
on average, for a given dark matter halo mass, twice more than with
our ejected-rate ($m_2$). We see the same trend if we compare the
two models at a fixed SFR.\\

The ejected-hot gas is transferred to the hot-halo phase. Its
possible definitive ejection from the hot atmosphere is computed by
assuming the dark-matter potential well, taking the
velocity of the wind and the escape velocity of the dark matter
structure into account. The gas mass that is definitively ejected is higher in
$m_2$ than in $m_3$, on average, in the intermediate range of masses.
This is linked to the wind velocity that is fixed to a value of about
$\simeq 100-150$~km/s in $m_2$ (see Eq.~13 in
\citealt{Somerville_2008}) and more than $\sim$150 km/s, on average,
in $m_3$\footnote{For a given kinetic energy the larger the wind
velocity, the smaller the mass ejected mass.}. This difference is at
the origin of the break in the slope of the stellar-mass function
between $m_2$ and $m_3$.

The large difference, at high mass, between $m_2$ and $m_3$ is due to
the AGN feedback. For consistency reasons with our hot gas phase
heating modelling (which associates both SN and AGN), we cannot apply
the AGN-feedback processes in this \cite{Somerville_2008} model
comparison, while, in model $m_2$, our AGN feedback is turned on, and
therefore reduces the stellar mass.\\

Despite the different parameterizations and energy injection scales
for supernovae, currently the classical semi-analytical models do not
seem to be able to explain the high-redshift behaviour of the mass
function in the low-mass range (see also Fig. 23 in \cite{Guo_2011},
and Fig. 11 in \cite{Ilbert_2013}). Even if some SAMs, such as
\cite{Somerville_2008}, \cite{Guo_2011}, or \cite{Henriques_2013}, use
a dedicated parametrization to reproduce the galaxy properties at
$z=0$, it seems that, at high redshift, the low-mass range problem of
the stellar-mass function is not only linked to a SN-feedback
efficiency calibration. Indeed, \cite{Guo_2011} (their Figs. 8 and 23) show
that the number of low-mass star-forming galaxies are still larger
than observed. A new ad-hoc parametrization of the
  \cite{Guo_2011} model is proposed
  by \cite{Henriques_2013}. Using a very high efficiency for the SN
feedback coupled to a very low efficiency for the re-accretion of
the gas (see Sect.~\ref{munich_model} for a complete discussion), they
obtain a better result in the low-mass regime.\\ 

A strong increase in the SN-wind efficiency
in low-mass structures also leads to very high mass-loading factors
($\dot{M}_{ej}/\dot{M}_{\star}>10$, \cite{Henriques_2013} their Fig.
3). Such factors are much greater that those derived from
spectroscopic observations \citep[e.g.][]{Sturm_2011, Rubin_2011,
Bouche_2012} even if the measurement of this parameter is
difficult and is currently performed on massive systems. In these conditions, stellar outflows alone cannot
limit the star formation sufficiently. In this context, the measurement of the
mass-loading factor becomes a key point. 

\section{An ad hoc recipe for reconciling models and observations}
\label{ad-hoc}

At high redshift ($z > 1$), as shown in
Fig.~\ref{mass_function}, the amplitude of the faint end of the
stellar mass function is dramatically over-estimated by the models
($m_1$, $m_2$ and $m_3$). This result is consistent with the
overestimate of stellar mass in low-mass dark matter haloes: small
structures form too many stars. In general, this problem is addressed
by a strong SN feedback and/or photoionization. As shown previously,
photoionization and SN-feedback cannot be sufficient to reduce
significantly the star formation in low-mass objects. Strong feedback
models give some good integrated results (at $z \simeq 0$)
\citep{Guo_2011, DeLucia_2007} but fail at higher redshift (see
\citealt{Ilbert_2013}, their Fig. 11).

In this section we propose a strong modification of
implementing of the star-formation mechanism in our semi-analytical
model to try to reconcile models and observations.

\subsection{Can all the cold gas form stars?}

In a standard semi-analytical model, the SFR
is adjusted to follow the observed empirical \cite{Kennicutt_1998}
law. The rate is computed following Eq.~\ref{star_formation_law} and
is applied to the entire cold gas reservoir. In this context,
the efficiency parameter $\varepsilon_{\star}$ determines the fraction
of star-forming gas. This fraction is obviously constant. The
\cite{Kennicutt_1998} law reflects, with global variables, an overall
view of the star formation process. Even if large reservoirs of gas
are observed in galaxies, at least up to $z\simeq1.5$
\citep[e.g.][]{Daddi_2010}, these observations do not give any
information about the real fraction of the gas that is available to
form stars.

The Kennicutt law and the homogeneous description of the
cold gas cannot describe the complex structure of the ISM.
Observations indicate that only a very small amount of the gas mass
is used at a given time to form stars in galaxies (including ours);
the star formation occurs only in highly-concentrated regions and not
in the entire disc. Recent Herschel observations show that stars form
in dense cold cores with a typical size of $0.1~pc$, embedded in the
interstellar filamentary structure \citep{Andre_2010, Heiderman_2010,
Lada_2012}. These prestellar cores are formed only when the gas
surface density is higher than $\Sigma_{thr}\simeq 160~\Msun\cdot
yr^{-1}$. Observations show that only a small amount of the total
gas mass ($\simeq 15\%$) is above this column density threshold and
only a small fraction ($\simeq 15\%$) of this dense gas is in the
form of prestellar cores \citep{Andre_2013a, Andre_2013b}. Therefore,
a large amount of the gas is not available to form stars. This
no-star-forming component corresponds to the gas that is
occupying the low levels of matter structuration, where the gas
surface density is low.

\subsection{The no-star-forming disc component} 
\label{ad-hoc-2}

When accreted on the galaxy disc, the surface density of
fresh gas (considered as homogeneously distributed) is low.
Progressively the gas, controlled by the turbulence and gravity
energy balance, is structured more and more \citep{Kritsuk_2011b}.
The energy injected by the accretion process must be dissipated
before star-formation process can start. Since the dissipation scale is
much smaller than the energy injection scale, we assume that the
energy cascade introduces a delay between the accretion time and the
star formation time.

To model this process, we introduce a model $m_4$ with a new
gas component in galaxy discs: the no-star-forming gas. The
delay between the accretion time of fresh gas and the time when this
gas is converted into stars is modelled by a transfer rate between
the no-star-forming gas and the star-forming gas
reservoir ($\sfg$) that follows
\begin{equation}
    \dot{M}_{\sfg,in} = \dot{M}_{\nosfg,out} =
\varepsilon_{\star}min\left[1,\left(\dfrac{M_{h}}{10^{12}~\Msun}\right)^3\right]\dfrac{M_{\nosfg}}{t_{dyn}}
    \label{no-sfg2sfg}
\end{equation}
where $M_{\nosfg}$ is the mass of no-star-forming
gas, $t_{dyn}$ the disc dynamical time, and $\varepsilon_{\star}$
an efficiency parameter, identical to the star formation efficiency
(Eq. \ref {star_formation_law}). Obviously this formulation is
totally ad hoc and does not provide any physical information on the
link between the characteristic time of the turbulent cascade and the
mass of the halo. The halo mass dependence in Eq.~\ref{no-sfg2sfg} is
introduced to reproduce the shape of the stellar to halo mass
relation, as observed in for example \citealt{Leauthaud_2012, Moster_2010,
Behroozi_2010, Bethermin_2012a}.This formulation has no other
purpose. It does not describe the structuration of the density.
However, the dependence in $M_h^3$ indicates that the star formation
regulation process must be extremely strong in the smallest
structures. In the context of the bimodal accretion, the accretion is
dominated by the cold mode below $M_h = 10^{12}\Msun$. This
cold accretion is feeding the no-star-forming gas reservoir,
which thus regulates the star formation in such structures.

In the context of this new prescription we decided to apply
the merger boost factor (Eq.~\ref{boost_factor}) to the
no-star-forming transfer process (Eq.~\ref{no-sfg2sfg}) and
to the star formation. Indeed we consider that mergers increase
 the mean gas concentration instantaneously, and thus accelerate the
structuration of the density.

\begin{figure}[t]
\begin{center}
 \includegraphics[scale = 0.3]{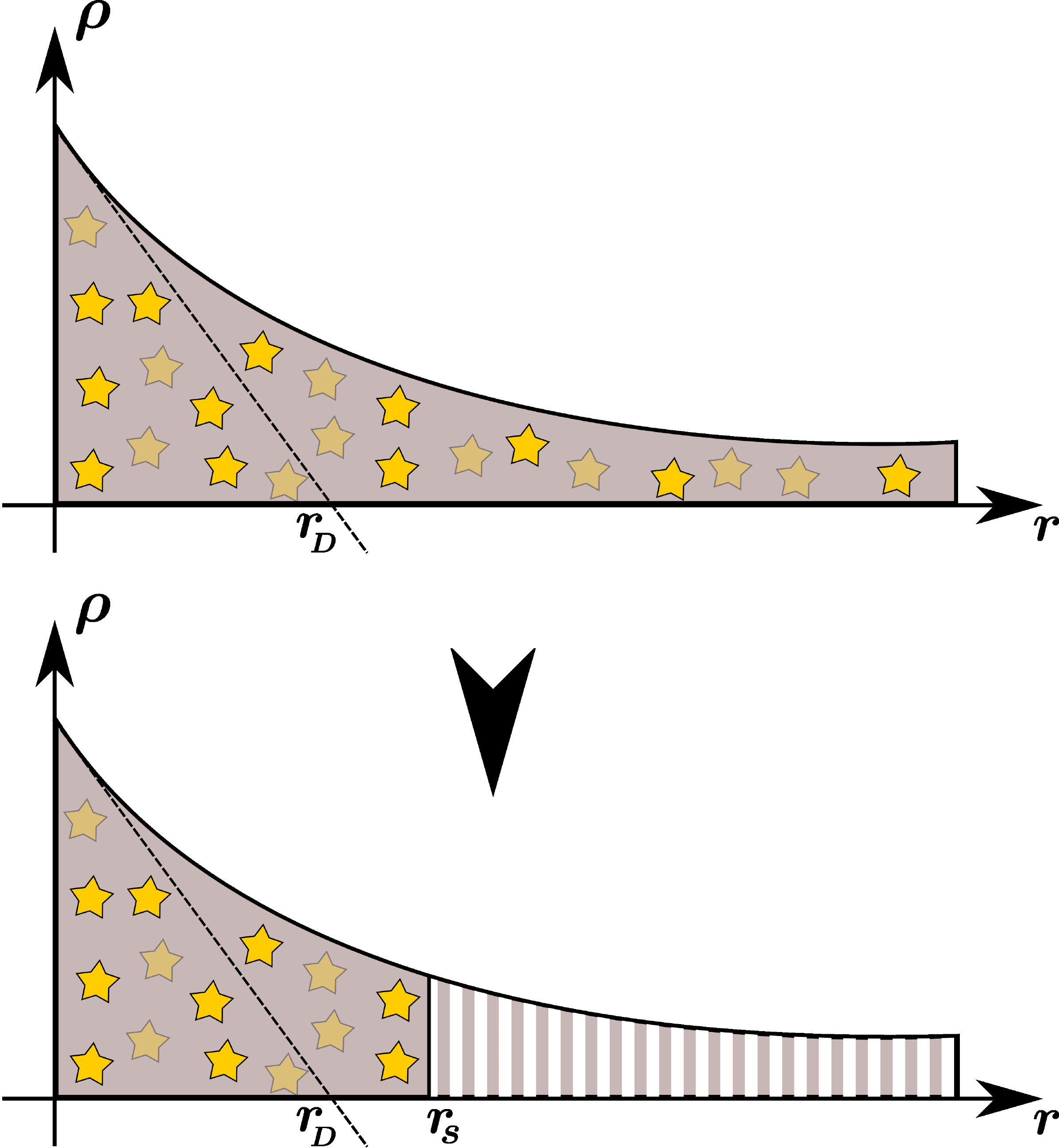}
 \caption{\tiny{Radial density profile of the gas in the disc. While
in a classical model (upper panel) all the gas is available to form
stars and is distributed in the whole disc, the star-forming
gas is artificially concentrated in the centre of the disc and the
no-star-forming is distributed in the outer region with our
ad hoc model (lower panel). The star-forming gas is enclosed
in the radius $r_s$.}}
  \label{no_sfg_distribution}
  \end{center}
\end{figure}

If the no-star-forming gas was homogeneously added to the
disc structure, the decrease in the star-forming gas
fraction would be equivalent to a simple decrease in the star
formation efficiency $\varepsilon_{\star}$. This is not satisfactory,
and to maintain the star formation efficiency even with a large
amount of no-star-forming gas, we thus had to adopt an
artificial gas concentration, as illustrated in Fig.
~\ref{no_sfg_distribution}. Using this gas redistribution, we derive
a new dynamical time, and thus SFR,  from the circular velocity
computed at the characteristic radius $r_s$. With this model, we can
produce high SFR, even if a large amount of gas is considered as
no-star-forming, without modifying the star formation
recipes.

\begin{figure*}[!t]
  \begin{center}
    \includegraphics[scale =
0.78]{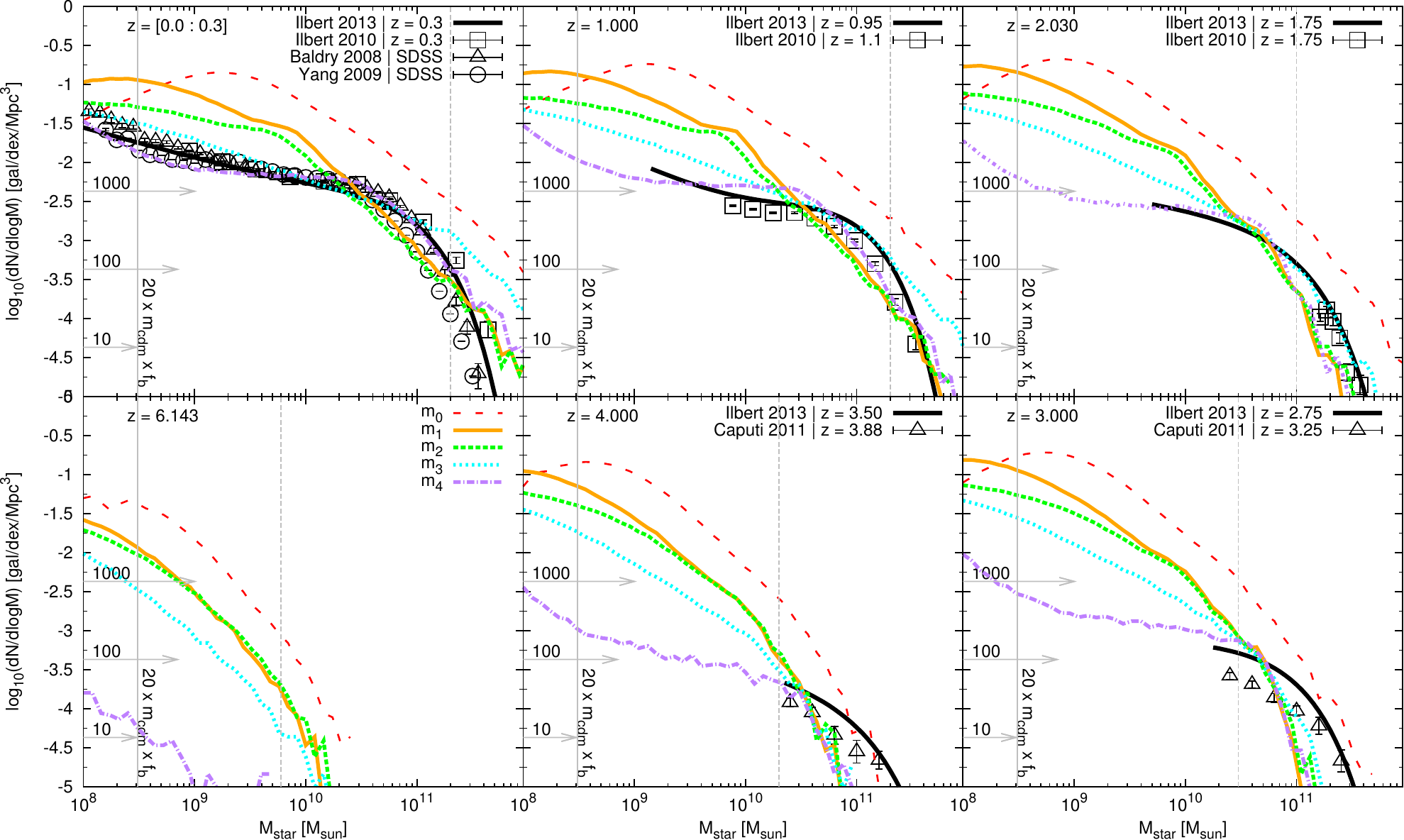}
  \caption{\tiny{Stellar mass function and its evolution with
redshift. The redshift is increasing clockwise. The colour code is
the same for all figures, and is detailed in the model list (Table
1). We compare our results with \cite{Ilbert_2010, Ilbert_2013}
(squares),  \cite{Yang_2009} (circles), \cite{Baldry_2008} (triangles
in the first panel), and \cite{Caputi_2011} (triangles) observations.
Horizontal arrows show the link between the density and the number of
haloes in our simulation volume. The grey dashed lines plotted in the
high-mass range indicate the limit where uncertainties due to the
cosmic variance are equal to the differences between models.}}
  \label{mass_function}
  \includegraphics[scale =
0.78]{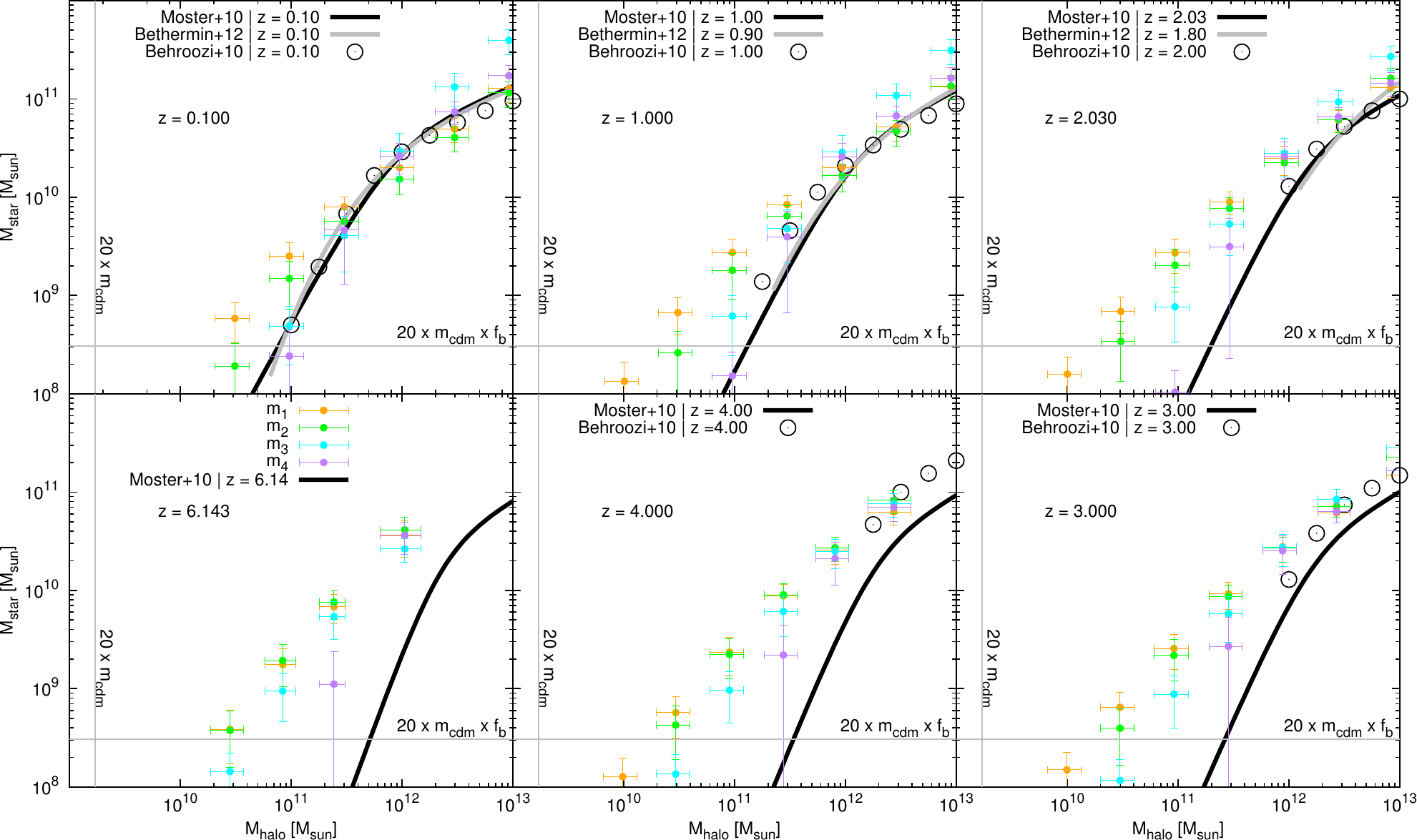}
  \caption{\tiny{Stellar-to-dark-matter halo mass relation (SHMR) for
various redshifts. The models (coloured lines) are compared with
recent analysis based on halo occupation or abundance matching (black
line: \cite{Moster_2010}; grey line: \cite{Bethermin_2012a}; black
open circles: \cite{Behroozi_2010}). While standard models have a
regular evolution of the stellar mass with $M_h$ , the ad hoc recipe
included in $m4$ produces: i) a very slow increase of the stellar
mass for haloes with $M_h < 10^{11}~\Msun$; ii) a very strong increase
of the stellar mass at intermediate halo mass ($10^{11} < M_h <
10^{12}~\Msun$). This shape is not captured well by our
points with error bars. The large error bar associated with the $M_h
= 10^{11}~\Msun$ point results from the very large scatter produced
by the strong increase of the stellar mass in this halo mass range.
iii) A slow increase in the stellar mass for $M_h > 10^{12}~\Msun$.}}
  \label{Mhalo_Mstars}
  \end{center}
\end{figure*}

\subsection{Comparison with the ``Munich model''}
\label{munich_model}

The current baseline of the ``Munich model'' is mainly
described in \cite{Guo_2011}. Some important modifications are
presented in \cite{Henriques_2013}. This paper focusses on the
reincorporation of the ejected gas. The model is based on an ejecta
reservoir that receives the gas ejected from the galaxy. The main
hypothesis is that this gas is not available for cooling and it has
to be reincorporated into the hot gas reservoir to cool. This ejecta
reservoir, linked to the halo, is fed by the very efficient
SN-feedback processes. As presented in \cite{Henriques_2013} the key
point of this model is the reincorporation timescale. In the current
model, it is inversely proportional to the halo mass and is
independent of redshift (from $1.8\times 10^{10}~yr$ for haloes with
$M_h = 10^{10}~\Msun$ to $1.8\times 10^{8}~yr$ for haloes with $M_h =
10^{12}~\Msun$). In this context, the gas expelled from low-mass
structures is stored for a long time in the ejecta reservoir, so this model strongly limits the star formation process. In
this scenario the no-star-forming gas is stored outside of
the galaxy. This model gives good predictions in the low-mass range
of the stellar mass function, but, without a prompt reincorporation
of this gas in the cooling loop, the amount of star-forming
gas, and therefore the star formation activity, are limited in the disc of low and intermediate-mass objects.

Regardless the mechanism behind it, it seems that the storage of
the gas in a no-star-forming reservoir (e.g. reservoir of
low-density gas in the disc, or reservoir without any cooling outside
the galaxy) is the best way to modulate the star-formation efficiency
such that semi-analytical models can reproduce the observations.

\section{Stellar and gas-mass assembly}
\label{discussion}

\begin{figure*}[t]
\begin{center}
 \includegraphics[scale =
0.72]{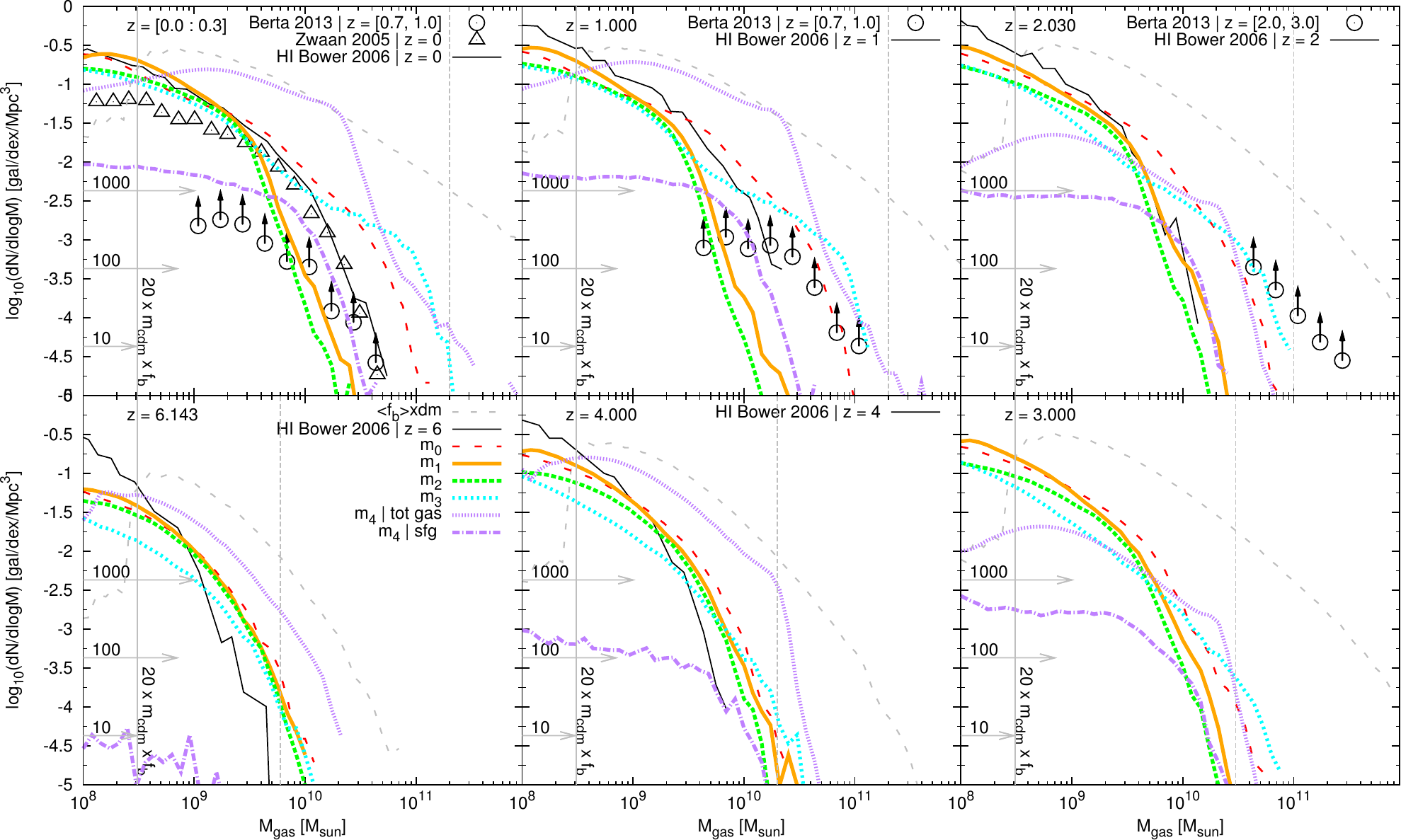}
  \caption{\tiny{Gas mass functions predicted by our SAMs. The colour
code is explained in Table \ref{model_description}. For comparison, we
add the extremal gas mass function deduced from the dark-matter mass
function and the universal baryonic fraction $\left<f_b\right>$ (grey
dashed-line). In the case of $m_4$, we plot the total
(star-forming + no-star-forming) and the
star-forming gas mass function. We compare our results with
the molecular gas mass function computed by \cite{Berta_2013} (lower
limits, circles) and with the local HI mass function computed by
\cite{Zwaan_2005}  (triangles). The black solid line shows the HI
mass function predicted by \cite{Lagos_2011}, using \cite{Bower_2006}
SAM. The horizontal arrows show the link between the density and the
number of haloes in our simulation volume.}}
  \label{gas_mass_function}
\end{center}
\end{figure*}

We have presented different processes that act on galaxy formation
and, more precisely, on the star formation activity. We tested two
photoionization models and two supernovae feedback models. In
addition to these four models, we have proposed another model in
which we have strongly limited the star formation efficiency in
low-mass haloes. In this section we discuss the comparison of this
model with the main galaxy properties, and compare the
predictions of this model in detail with the other four.

\subsection{Stellar-mass function and SHMR}

We show in Fig.~\ref{mass_function} the stellar-mass functions
predicted by our models. Model outputs are compared with
observational data from \cite{Ilbert_2010}, \cite{Ilbert_2013},
\cite{Baldry_2008}, and \cite{Yang_2009,Caputi_2011}. \\ 
As discussed before, models $m_1$, $m_2$, and $m_3$ fail to reproduce
the low-mass end of the stellar mass function. The disagreement is
both on the amplitude (one order of magnitude higher at low mass) and
on the shape of the mass function. Figure~\ref{Mhalo_Mstars} shows
the stellar mass as a function of the dark-matter halo mass. We
compare all models with relations extracted from the literature
\citep{Behroozi_2010,Bethermin_2012a,Moster_2010}. This figure
indicates that the excess of galaxies with low stellar masses is due
to an over-production of stars in the low-mass dark matter haloes. To
reduce this tension, we applied a strong modification of the
star formation process in $m_4$. The gas is kept in the disc but, a large
amount of this gas cannot form stars. With this ad hoc model, in the
low mass range, the levels of the stellar mass functions are in good
agreement with observations for a wide range of stellar masses. This
indicates that only a strong modification (a decrease in our case) of
the mass of gas instantaneously available to form stars allows the star formation activity in low mass structures to be
modulated and SAM to be
reconciled with the observations.

Concerning the high-mass end of the stellar-mass function, all models
under-predict the number of massive galaxies. For $z = 4$ and $z=3$,
the comparison with \cite{Ilbert_2010} and \cite{Ilbert_2013}
observational mass functions indicate that the massive galaxies in
our models are two time less massive than the observed distribution.
This is also observed in other recent SAMs (see e.g. \citealt[their
Figs. 4, 5 and 6]{Henriques_2013}, and \citealt[their Fig. 23]{Guo_2011}).
The only way to reconcile models and observation in this high-mass
regime is to consider a model without any regulation mechanism (model
$m_0$). In contrast, for the low-redshift range ($z = 0-2$), our
models give a small excess for massive galaxies. This disagreement
could be linked to an AGN-feedback that is not efficient enough.
The volume in the simulated box
[$(100/h)^3\simeq 150^3~Mpc^3$] does not allow us to have more than
ten haloes with mass higher than $M_h = 10^{13}~\Msun$, and therefore
there is a small statistical sample associated to this range of mass. As we can see in Fig
\ref{mass_function} at $z = 0.3$ and for mass larger than
$10^{11}~\Msun$, the stellar mass function is quite noisy. For
information, we indicate, in each panel, the stellar mass above which uncertainties due to cosmic
variance become larger than the differences between the models (see
Appendix \ref{cosmic_variance} for more details).

\subsection{Gas-mass function}

In $m_4$, we have chosen to modify the standard star
formation paradigm, through the introduction of a delay between gas accretion and
star formation. The step during which the no-star-forming
gas is converted into star-forming gas strongly reduces the
star formation activity, and therefore the stellar mass build-up.
Obviously the total amount of gas in galaxies
(no-star-forming and star-forming) will be strongly
modified with respect to standard models. In this section, we compare the
gas mass function predicted by all models with the available
observational constraints.

In Fig.~\ref{gas_mass_function}, we show the
predicted gas-mass functions, together with the local HI mass function computed by
\cite{Zwaan_2005} and the molecular gas mass function coming
from \cite{Berta_2013}. In their study, \cite{Berta_2013} focus on the molecular gas contained in normal star-forming galaxies (within $\pm~0.5$ dex in SFR from the main
sequence). Quiescent galaxies are therefore not taken into account.
In these conditions, and as explained by \cite{Berta_2013}, their
data points should be considered as lower limits. The gas-mass functions extracted from our models
are computed using all galaxies contained in our simulated volume,
and taking the total gas mass in galaxy discs into account 
(both star-forming and no-star-forming).

The gas-mass functions predicted by our reference model
$m_1$ and its variation ($m_2$) are very close. Indeed, the two
models use the same prescription for gas ejection (sn + agn). At low
mass and at all redshifts, the gas mass predicted by $m_3$ (using
\cite{Somerville_2008} sn feedback prescription) is also very close
to $m_1$ and $m_2$. At high mass, the difference is due to SMBH.
Indeed, in $m_3$ the SMBH activity is not taken into account.
Consequently, the cooling rate associated with high-mass haloes is
not limited, and the amount of gas increases.

In the case of the new model $m_4$, we plot in
Fig.~\ref{gas_mass_function} the total and the star-forming
gas-mass function. As expected, the amount of total gas in $m4$ is
larger than in the reference model $m_1$ or its variations ($m_2$ and
$m_3$). The decrease in the star formation activity in $m_4$ leads to
an large storage of the gas. In the first two panels ($z\simeq 0$
and $z\simeq 1$), the difference between the total
and the star-forming gas mass functions is larger than at
higher redshift; the fraction of no-star-forming gas
increases with time. This evolution is linked to the transfer rate
between the no-star-forming gas 
and the star-forming gas reservoirs (Eq~\ref{no-sfg2sfg}).
Indeed, for a given mass of no-star-forming gas, the rate
increases with the dark-matter halo mass, but only up to
$M_h=10^{12}~\Msun$. Above this threshold the rate is constant for a given mass of
no-star-forming gas. Thus, for haloes
more massive than $M_h=10^{12}~\Msun$, the fraction of
no-star-forming is increasing.

We also show in Fig.~\ref{gas_mass_function} the HI-mass
fonction derived by \cite{Lagos_2011} using the SAM of
\cite{Bower_2006}. At first order, it is comparable to the
mass-function evolution from our reference model $m_1$, which is reassuring 
and expected. 
It is interesting to note that their predictions for the low-mass range ($M <
10^{9}~\Msun$) are systematically higher than those predicted by our
models at $z>0$. This effect could be linked to some resolution effects.
At $z=0$, the reference model $m_1$ under-predicts
the mass function at high mass, while the \cite{Lagos_2011} model shows better agreement with the measured HI mass function. This difference
can be due to a stronger SMBH feedback in our reference model. Under
this hypothesis, in $m_1$, with less SMBH feedback, the accretion
rate, hence the SFR, would be higher. This would
increase the assembled stellar mass and thus the level of stellar
mass function that is already too high in the high-mass range.\\

At $z\simeq 0$, the amount of total gas predicted by $m_4$ is
larger than the measurement of the HI gas and the lower values of 
the molecular gas. Independently of each
other, the HI and the molecular gas represent only a fraction of the
total gas mass contained in a galaxy. However, we can note that for
the high-mass range, the star-forming gas mass function
predicted by $m_4$ is in good agreement with the HI mass function
measured by \cite{Zwaan_2005}. Even if the total gas mass predicted
by $m_4$ seems high, without any measurement of this total mass, it
is difficult to conclude. The total gas mass function appears today
as one of the key observables that will allow us to determine the
optimal efficiency of gas ejection process and star formation.

We can integrate at $z=0$ the HI-mass function measured by
\cite{Zwaan_2005} and the gas-mass function predicted by model $m_4$
in the mass range $[10^8,10^{12}]~\Msun$ to compare the gas mass
fractions in the different components
(Table~\ref{gas_mass_ratios}). If we gather the observed and the 
predicted values, these ratios indicate that
\begin{itemize}
  \item{only $\simeq$ 2.5\% of the total gas mass contained in a
galaxy can be used directly to form a new generation of stars;}
  \item{$\simeq$ 70\% of the HI mass should be no-star-forming,} 
  \item{more than 90\% of the no-star-forming gas is not
detected in HI. This fraction is potentially 
greater than observed, even if its nature still has to be
examined in the context of the ``missing baryon problem''. 
It could be low-metallicity molecular gas or very hot (X-ray)
diluted gas. For instance, some years ago \cite{Pfenniger_Combes_1994a} and \cite{Pfenniger_Combes_1994b} proposed that there is a large amount of cold \textit{dark} gas
(essentially in molecular form, H2 in a fractal structure) evolving 
in the outer parts of galaxy discs. Following
\cite{Pfenniger_Combes_1994b}, this gas could be
no-star-forming and in equilibrium between
coalescence, fragmentation, and disruption along a hierarchy of
turbulent clumps. In these conditions, the dissipation time of this
gas may exceed several Gyr, and only a tiny fraction may be turned
into stars. This kind of hidden baryon could be a candidate for a 
physical explanation of the ad hoc model. As already explained,
the total gas mass function appears as a key observable in the context
of the new molecular gas surveys.} 
\end{itemize}

Comparing models and observations at higher redshift, we see from
Fig.~\ref{gas_mass_function} that standard models strongly
under-predict the amount of observed (molecular) gas, at least for $z\le2$ in high-mass
galaxies. At high masses (dark-matter halo + galaxy), the gas-ejection
process is very limited. Thus, the lack of gas mass cannot be
explained by an excessive instantaneous ejection in high-mass
objects, but by a time-integrated ejection process that is too
efficient.

At $z\sim2$ the amount of molecular gas measured by
\cite{Berta_2013} is much greater than predicted by all models.
Even if the observed gas quantity seems very large, the lack of gas
in the models may explain the under-prediction of massive galaxies at
these epochs (see Fig.~\ref{mass_function}).

\begin{table*}[t]
  \begin{center}
    \footnotesize{
      \begin{tabular}{cccccc}
        \hline
        HI/gas & SFG/gas & noSFG/gas &  (HI$-$SFG)/HI &
(noSFG$-$HI)/noSFG & noSFG/bar\\
        (a) & (b) & (c) & (d) & (e) & (f)\\
        \hline
        0.084 & 0.025 & 0.975 & 0.70 & 0.91 & 0.65\\
        \hline
      \end{tabular}}
  \end{center}  
  \caption{\footnotesize{Gas-mass fractions at $z=0$ of (a) HI
gas w.r.t total gas in galaxy (gas), (b) star-forming gas (SFG) w.r.t. total
gas in galaxy, (c) no-star-forming HI gas (HI$-$SFG) w.r.t. HI, (e) no
star-forming \& no HI gas w.r.t. no star-forming gas, (f) no-star-forming gas w.r.t baryon in the halo}}
  \label{gas_mass_ratios}
\end{table*}

\subsection{Star formation rate history}
\label{star_formation_history}

\begin{figure}[]
    \centering \includegraphics[scale = 0.78]{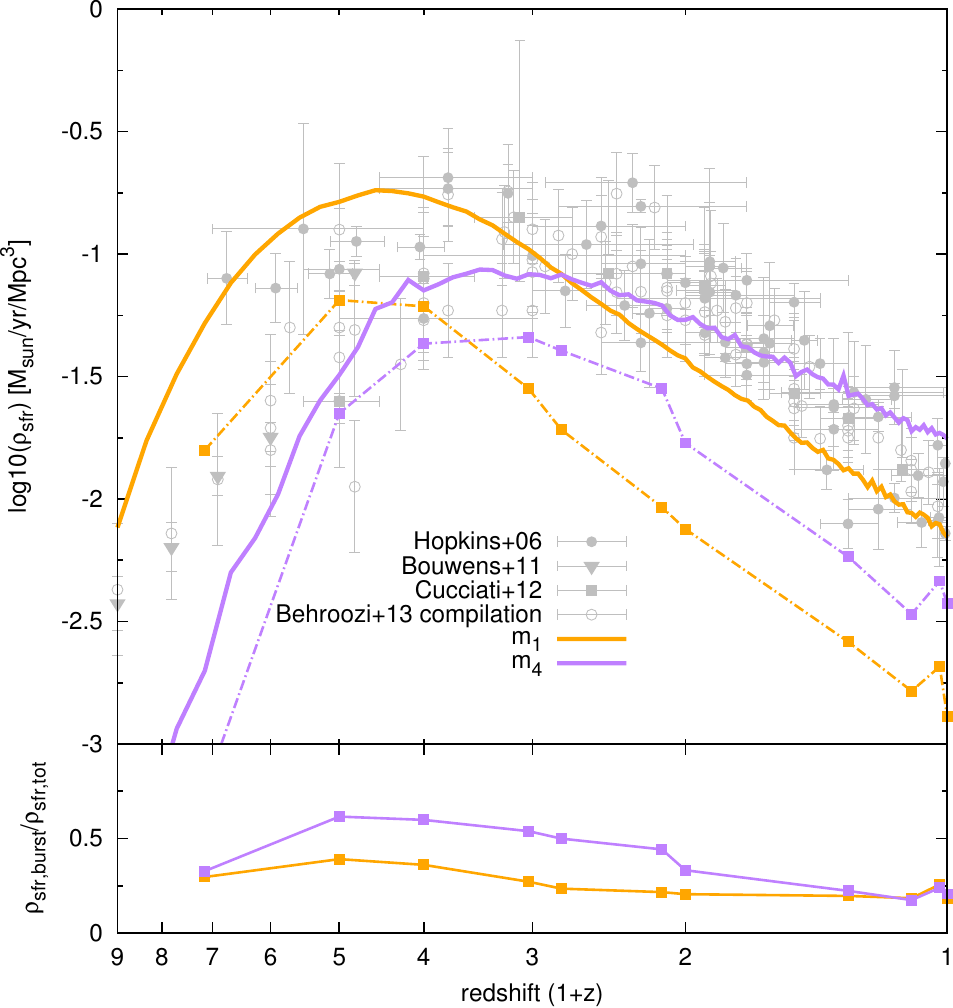}
  \caption{\tiny{Redshift evolution of the star formation rate
density. Model predictions (coloured lines) are compared with
observational data (grey points) coming from \cite{Hopkins_2006},
\cite{Bouwens_2011}, and \cite{Cucciati_2012}. Solid lines present the
star formation rates derived from the models. They are computed as
the sum of the star formation rate of all galaxies and divided by the
box volume [$(100/h)^3\simeq 150^3~Mpc^3$]. Dashed lines show the
merger-driven star formation activity. The bottom panel
gives the redshift evolution of the ratio
$\rho_{sfr,burst}/\rho_{sfr,tot}$. We can see that the fraction of
star formation activity linked to merger events is larger in $m_4$
than in the reference model $m_1$.}}
  \label{Madau_plot}
  \centering \includegraphics[scale = 0.78]{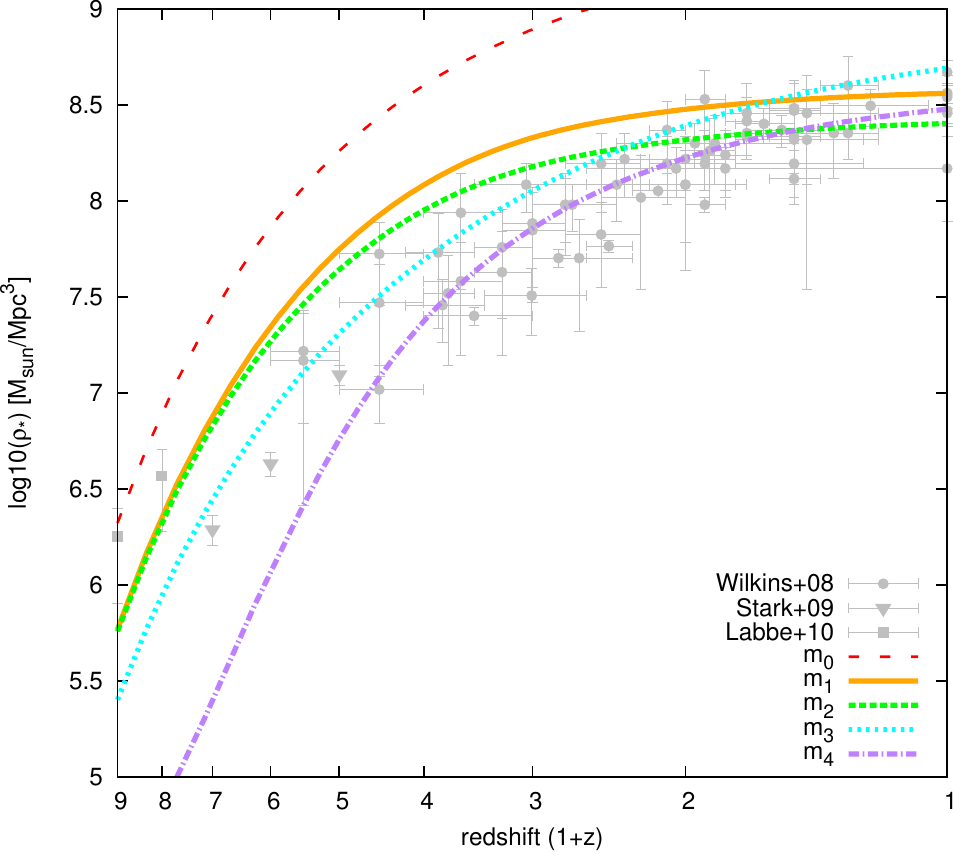}
  \caption{\tiny{Redshift evolution of the stellar mass density from
the models (coloured lines). Model predictions are compared with
observational data (grey points) coming from \cite{Wilkins_2008},
\cite{Stark_2009}, and \cite{Labbe_2010}. }}
  \label{rho_star}
\end{figure}

Figures \ref{Madau_plot} and \ref{rho_star} show the evolution with
redshift of the cosmic SFR density (CSFRD) and the
stellar mass density, respectively. Models are compared to a set of
observational data: \cite{Hopkins_2006}, \cite{Bouwens_2011}, and
\cite{Cucciati_2012} for the SFR density, and
\cite{Wilkins_2008}, \cite{Stark_2009}, and \cite{Labbe_2010} for the
stellar mass density. For the sake of clarity, only $m_1$ and $m_4$ are
shown in Fig.~\ref{Madau_plot}. Models $m_2$ and $m_3$ lead to
similar evolutions to model $m_1$.\\

Model $m_1$, with (SN/AGN)-feedback, presents a peak of CSFRD between
$z = 3$ and $z = 6$ that is marginally compatible with observations.
This early peak in star formation is at the origin of the
over-production of stellar mass (Fig. ~\ref{rho_star}) in the
structures formed at this epoch (with $M_{halo} \le 10^{11}~\Msun$).
A strong SFR leads to a strong SN feedback,
and consequently to a large amount of mass that definitively leaves
the dark matter halo potential ($\simeq 60\%$). The gas density in
the hot atmosphere becomes too low to produce efficient cooling,
and the accretion rates for the galaxies decrease. For example, with
our model $m_1$ at $z=2$, the mean cooling rate on a
$10^{11.5}~\Msun$ dark matter halo is $2.15~\Msun\cdot yr^{-1}$. The
cooling rate falls at $1.40~\Msun\cdot yr^{-1}$ at $z=0.3$, which is
a decrease of 35\%. This lack of fresh gas is at the origin of the
strong decrease in the SFR found for $m_1$ at low
redshifts ($z < 3$). Even if the boost factor
(Eq.~\ref{boost_factor}) is applied to the post-merger galaxies,
$m_1$ cannot correct for this lack of star formation. In the ``Munich
model'' this decrease is compensated by the reincorporation of the
gas previously expelled from the galaxy on a nadapted timescale,
which allows increasing the gas mass available to cool.\\

The redshift evolutions of the CSFRD and stellar mass
density predicted by $m_4$ are in better agreement with observations.
Indeed, the strong reduction of the gas fraction available to form
stars allows both reducing the star formation activity at high $z$
and maintaining a larger amount of gas than in model $m_1$ at low $z$.\\ 

Fig.~\ref{Madau_plot} shows, the solid lines  the total CSFRD. In
the same plot we added a second measurement for information, where we have only taken the merger-driven star-forming
galaxies. This kind of galaxy is defined as a post merger structure
with a SFR boost (Eq.~\ref{boost_factor}) leading to SFR values greater
than $\mean{SFR}+\sigma_{SFR}$\footnote{$\mean{SFR}$ and
$\sigma_{SFR}$ are respectively the average and the standard
deviation of SFR distribution of steady-state
galaxies with stellar mass and dark matter halo mass value close (in
the same half dex) to the merger-driven starburst galaxy that was considered}. In the steady-state galaxies, the SFR is
closely linked to the fresh accretion of gas and is not due to recent
merger events. We see from the figure that at high redshift
($z\sim3$), in $m_1$, the stellar mass growth is dominated by the
star formation in steady-state objects. As explained previously, in
$m_4$ the merger driven star formation process is more efficient than
in $m_1$. Indeed, even if we apply the merger-boosting factor
(Eq.\ref{boost_factor}) to the star formation of post-merger galaxies
in $m_1$,  they do not have enough gas to produce large starbursts. \\

Fig.~\ref{rho_star} gives a good summary of the situation.
We see that for standard models ($m_1$, $m_2$ and $m_3$), the stellar
mass is formed too early in the evolution histories of galaxies.
Currently, standard SAMs \citep[e.g.][]{Somerville_2008, Guo_2011}
are reproducing the local CSFRD and stellar-mass density quite well.
But these good results at z=0 must not obscure the fact that the
history of the stellar mass assembly is not described well. Indeed,
as shown in Fig.~\ref{rho_star}, the stellar mass densities predicted
at $z>3$ are systematically larger than the observations. The results
obtained at z = 0 are only due to an excessive decrease in the mean
SFR after the strong over-production of stars at high
redshift. Indeed, as shown in Fig.~\ref{Madau_plot}, for $z < 2$ the
CSFRD is systematically lower than observed. This decrease is
also visible in the slope of the stellar mass density. If the gas is
over-consumed at high $z$, it seems to be missing at low $z$. Only
the ad hoc model that maintains a large amount of gas in discs
allows reproducing at the same time i) the stellar mass density at
all redshifts and ii) the trend in the CSFRD.

\subsubsection{The star formation rate distribution}

\begin{figure*}[]
  \begin{center}
    \includegraphics[scale =
0.8]{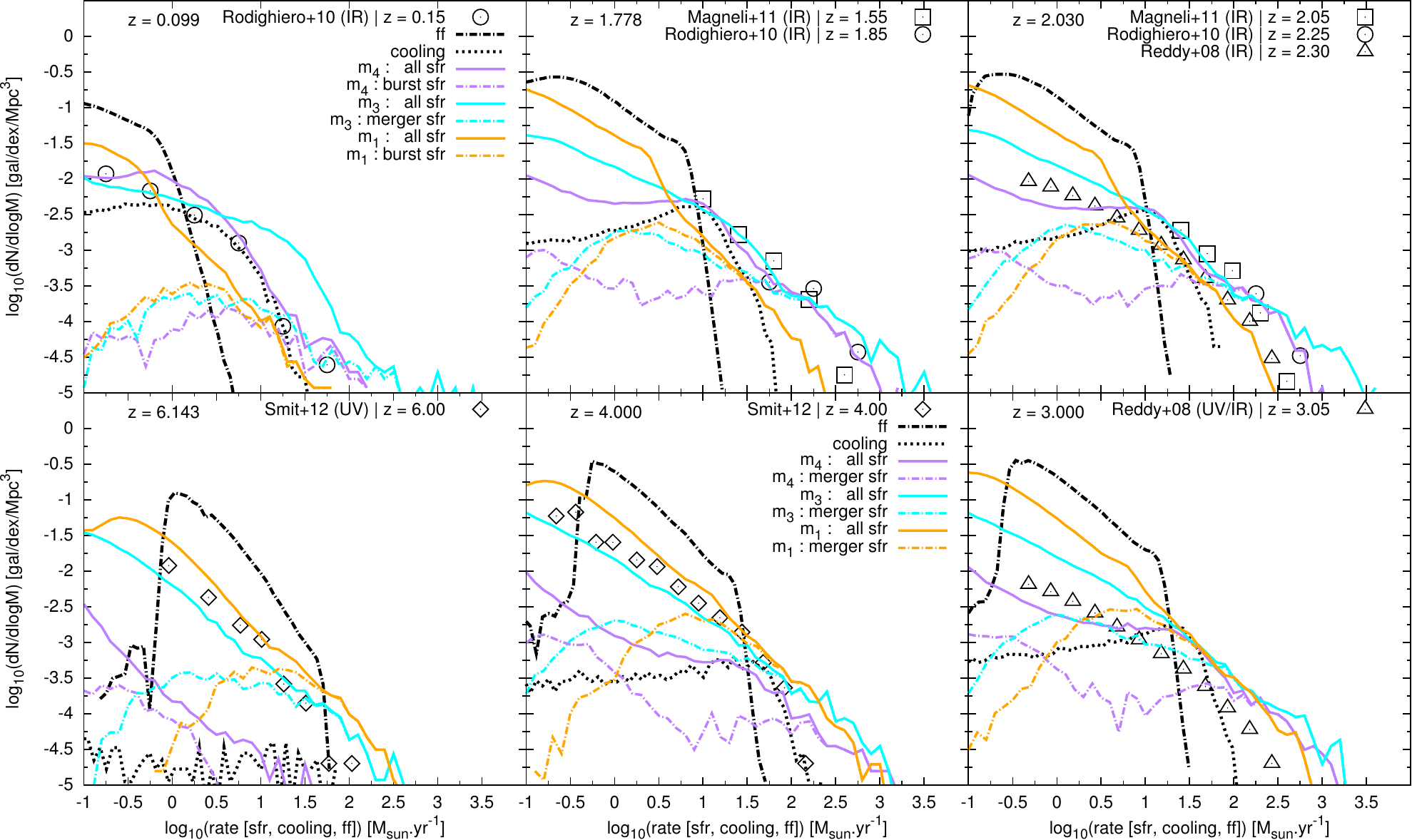}
  \caption{\tiny{Star formation rate distribution for $m_1$, $m_3$,
and $m_4$. Solid and dot-dashed coloured lines give the total and the
merger-induced (Eq.~\ref{boost_factor}) star formation rate
distribution, respectively. The merger-induced distributions are
built with galaxies that  i) have merged during the last time-step
and ii) have $SFR>~\mean{SFR}+\sigma_{SFR}$, where $\mean{SFR}$ and
$\sigma_{SFR}$ are, respectively, the mean and standard deviation of
star formation rate distribution of steady-state galaxies with
stellar mass and dark matter halo mass value close (in the same half
dex) to the merger-driven starburst galaxy taken into account. Models
are compared with a set of observational data coming from
\cite{Rodighiero_2010a} (circles), \cite{Magnelli_2011} (squares) and
\cite{Reddy_2008} (triangles) at low redshift ($z\le3$), and from
\cite{Smit_2012} (diamonds) at high redshift ($z=4-6$). For
information we add the free-fall rate (black dash-dot line) and
cooling rate (black dot line) distribution function for model $m_4$.}}
  \label{sfr_function}
  \end{center}
\end{figure*}

Fig.~\ref{sfr_function} shows the SFR
distribution function. We compare models $m_1$, $m_3$, and $m_4$ with
a set of observational data. To obtain the SFR distribution we
applied the standard conversions from infrared luminosity or
ultraviolet magnitude measurements to SFR (\cite{Kennicutt_1998}
conversions, with a Chabrier IMF, $SFR~[\Msun\cdot yr^{-1}] =
3.1\times 10^{-10}~L_{IR(8-1000\mu m)}~[L_{\odot}]$). In
Fig.~\ref{sfr_function}, we are also showing the total and
merger-driven SFR. In all cases, the high SFR values
are mainly linked to merger events, even if the boost factor
generates a wide range of SFR.\\

As discussed previously, the lack of gas in galaxies formed in $m_1$
produces a lower CSFRD than observed at low redshift.
As a result, the SFR distribution predicted by $m_1$ at low
redshift ($0<z<1.8$ in the figure) is always lower than the
observations by a factor 0.3\,dex on average. For $m_4$, the
artificial gas concentration allows a large star
formation activity to be maintained even if the fraction of star-forming gas
is low. This delayed star formation model gives a good
match to the observation for $0<z<3$. But this good result must be
put inot context. Indeed the strong decrease applied to the star-formation
rate in $m_4$ leads to a very low level of star formation at high
redshift ($z \simeq 6$), as can be seen in Fig.~\ref{sfr_function}.
Even if it seems that we need to strongly reduce the star formation
activity if we want to reproduce galaxy properties at low redshift
($z<3$), the comparison with the \cite{Bouwens_2007} measurements
indicates that model $m_4$ clearly under-predicts the star formation
rate at these epochs. The observations show that, in some structures
at high redshift (as for Lyman break galaxies), stars are formed with very high efficiency. Also, modelling the high-redshift Lyman-Alpha
emitters, \cite{Garel_2012} found that a very high star-formation
efficiency is needed to reproduce the luminosity function.
Our model $m_4$ cannot produce these kinds of galaxies.\\

To reconcile the predicted star formation rate distribution with the
observed one, but still producing the same amount of stellar mass in
these objects at these epoch\footnote{Indeed, as shown in the stellar
mass function, the build-up of the stellar mass agrees with
the observed stellar mass function.}, we need to have the same
quantity of star-forming gas, but to reduce the dynamical
time of the star formation process. To do that we could associate the
star-forming gas component to denser regions, with smaller
characteristic sizes. 
This is expected to give star formation rates that would be
comparable to those predicted by $m_1$. The SFR distribution in $m_1$
is in good agreement with \cite{Bouwens_2007} at very high $z$, even
if the stellar mass produced in these low-mass haloes is higher than observed. In this model, the rhythm of star formation is good,
but the star formation acts on gas reservoirs that are too large.
The star formation activity produced by $m_3$ (based on
\cite{Somerville_2008} SN-feedback), compared to the reference model
$m_1$, under-predicts this high-redshift star-forming population.
This is probably due to the strong photoionization and SN-feedback
processes considered in this model.

\begin{figure*}[!t]
     \centering \includegraphics[scale =
0.8]{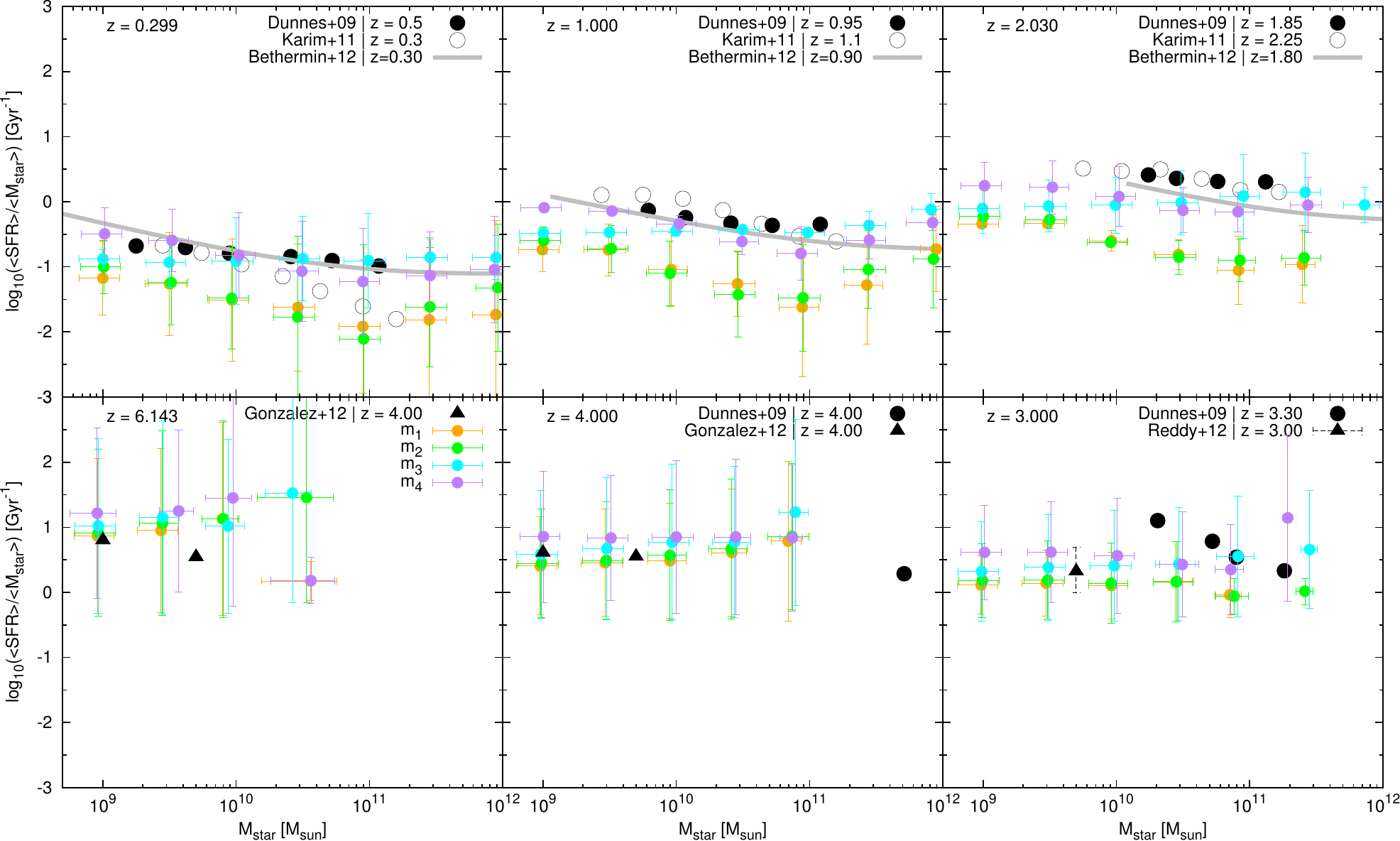}
  \caption{\tiny{Specific star formation rate as a function of the
stellar mass for a set of redshifts (the redshift is increasing
clockwise). This quantity can be seen as the inverse of the stellar
mass doubling time. The higher the value, the more intense the
star formation activity. We compare the models (colour-coded) with
observational data from \cite{Dunne_2009}, \cite{Karim_2011},
\cite{Gonzalez_2012}, \cite{Reddy_2012b}, and \cite{Bethermin_2012a}.
For most observational data points, errors are smaller than the size
of the symbol.}}
  \label{Mstars_ssfr}
\end{figure*}

Measurements of star formation rate, given by \cite{Reddy_2008} and
\cite{Bouwens_2007}, are computed using UV observations, corrected
from dust extinction. This correction is based on the UV
continuum/$\beta$-slope extinction law and is mainly an extrapolation
of results from local galaxies \citep[e.g.][]{Burgarella_2005}. Even
if this relation seems to be valid at redshift $z\simeq2$
\citep[e.g.][]{Daddi_2007, Reddy_2012b}, it has a huge scatter, and
extrapolations to large redshifts lead to large errors (especially at
high SFR) that are difficult to estimate. \\

For completeness, we added the
distribution function of the filamentary accretion rate to Fig. \ref{sfr_function} and the cooling rate for $m_4$. As
expected, the cold mode efficiency decreases when redshift decreases.
At high SFR, it is interesting to note that no galaxy is accreting enough baryons to form stars in a steady-state mode;
only the episodic merger events can produce these high values. 

\subsubsection{The specific star formation rate}

\begin{figure}[h]
  \includegraphics[scale = 0.8]{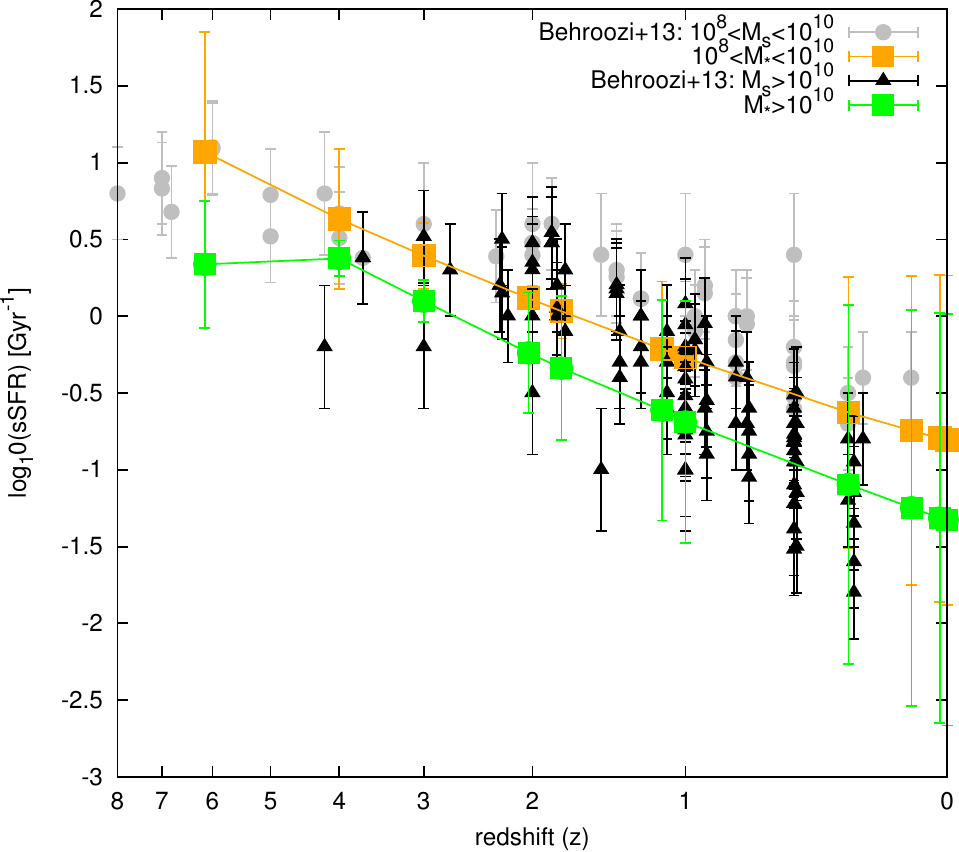}
  \caption{\tiny{Redshift evolution of the specific star formation
rate derived from $m_4$ ($10^8<M_{\star}<10^{10}\Msun$ in orange and
$10^{10}<M_{\star}<10^{12}\Msun$ in green). The error bars correspond
to the standard deviation. For comparison, we show the data points
around $M_{\star}=10^{9}\Msun$ and $M_{\star}=10^{10}\Msun$,
extracted from \cite{Behroozi_2013c}.}}
  \label{cosmic_SFR}
\end{figure}

\begin{figure*}[!t]
  \includegraphics[scale =
0.85]{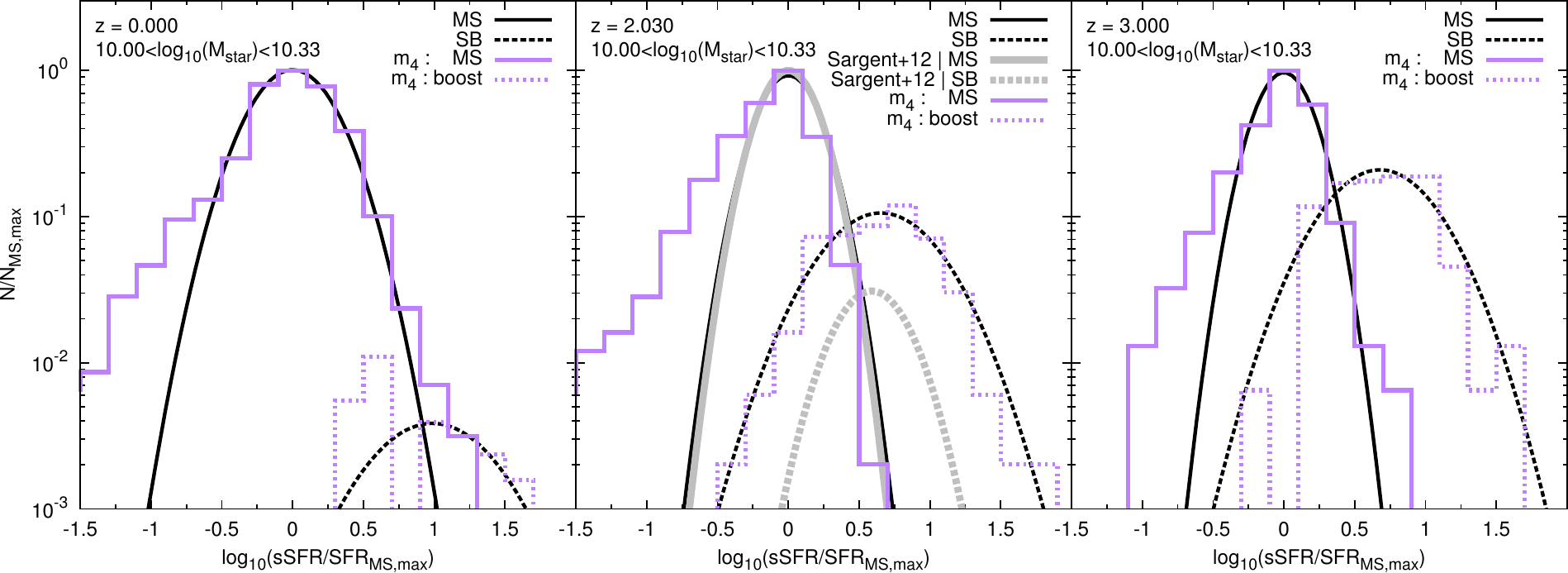}
  \caption{\tiny{Specific star formation rate (sSFR)
distributions derived from model $m_4$, for galaxies with a stellar
mass in the range: $10<log_{10}(M_{\star})<10.33$. The purple solid
histogram shows the distribution of galaxies in the steady state mode
(SFR$~\propto~$accretion rate). Such a distribution contains main
sequence (MS) and quiescent galaxies ($sSFR/SFR_{MS}<0.1$ in our
study). The purple dashed-line histogram shows the distribution of the
sSFR for galaxies with post-merger star-formation activity (PMSB)
(Eq. \ref{boost_factor} and following $SFR_{PMSB} \ge
\overline{SFR}+\sigma_{SFR}$). At $z=2$ we compare our results with
\cite{Sargent_2012} measurements in which the quiescent population
have been removed (grey solid line). For a better comparison with
these observational measurements, we add a
log-normal distribution as a black solid line adjusted only on our MS galaxy population. We
have removed all galaxies with $sSFR/SFR_{MS}<0.1$. We see that
model $m_4$ is in very good agreement with the MS distribution but
over-predicts the number of galaxy in the post-merger starburst mode
(0.4 dex).}}
  \label{sSFR_dist}
\end{figure*}

Fig~\ref{Mstars_ssfr} shows the specific star formation rate
($sSFR = SFR/M_{\star}$). We compare all models with a set of
observations from \citealt{Dunne_2009, Karim_2011, Gonzalez_2012,
Reddy_2012b} and with \cite{Bethermin_2012a} model predictions.
First, for all models we note that the mean sSFR (over the whole mass
range) increases with redshift. More specifically, we show in
Fig.~\ref{cosmic_SFR} the redshift evolution of the sSFR, extracted
from model $m_4$, for two mass ranges. We see that more massive
galaxies have lower sSFR at any redshift, implying that they form the
bulk of their stars earlier than their low-mass counterparts.  

At all redshift $z\le 3$ and for the whole mass range, models $m_1$
and $m_2$ are systematically lower (by a factor 10) than the
observations. This result is due, on the one hand, to the low SFR as
seen in Fig. \ref{sfr_function}, and on the other to the excess
of stellar mass as seen in Fig. \ref{Mhalo_Mstars}. \\

The model with stronger SN-feedback ($m_3$) gives better results in
the intermediate redshift and mass ranges ($0 < z < 3$ ;
$10^{10}<M_{\star}<10^{11}~\Msun$) but stays lower than the
observations in the lower mass regime for the same reasons as
explained previously. The predictions at high mass suffer from the
absence of AGN-feedback and do not have to be considered.\\

Model $m_4$ gives better agreement with the observations
than the reference model $m_1$ or its variation $m_2$ does. In $m_4$, the
mean star-formation efficiency is higher. This result does not
contradict with the main objective of this ad-hoc model.
The star formation activity is strongly reduced, and thus the
produced stellar mass is also strongly reduced. This two trends lead
to a higher level of the sSFR than in $m_1$ and $m2$.\\

Fig.~\ref{sSFR_dist} shows the sSFR distributions of
galaxies predicted by model $m_4$ in a limited stellar mass range
such that they can be directly compared with the \cite{Sargent_2012}
observational measurements. The population of quiescent galaxies
has been removed in \cite{Sargent_2012},
we added two log-normal distributions to our histograms to allow for
a better comparison:
\begin{itemize}
  \item{a first one fit the MS population;}
  \item{a second one fit the merger-driven population.}
\end{itemize}
The excess of objects at low sSFR in the histograms, in comparison to
the MS log-normal distribution, corresponds to the quiescent galaxy
population.
 
We can see that the MS distribution predicted by $m_4$ is in
excellent agreement with \cite{Sargent_2012}. However, the number of
galaxies in the PMSB mode is over-predicted even if the average
values of the distribution is in good agreement with what is observed
($log_{10}(sSFR/SFR_{MS})\simeq 0.6$). It is possible to reduce this
tension by modifying our PMSB definition ($SFR_{PMSB} \ge
\overline{SFR}+\sigma_{SFR}$). If we replace $\sigma_{SFR}$ by
$2\sigma_{SFR}$ in the previous definition, the number of objects on
the PMSB population obviously decreases and becomes comparable to
\cite{Sargent_2012}, but in this case, the centroid of the PMSB
distribution is then shifted to a higher value
($log_{10}(sSFR/SFR_{MS})\simeq 1$). The properties of the PMSB
population are also strongly dependent of the merger-boost factor
used in the model (Eq.~\ref{boost_factor}). Some modifications, such
as on the amplitude of the characteristic merger time scale
($\tau_{merger}$), will be explored in a future work.

\subsection{Steady-state versus merger-driven star formation}
To summarize, galaxies evolve in a quasi-steady state in $m_1$, $m_2$, and $m_3$, after the high
level of star formation activity at $4 < z < 6$, and consequently the
high rates of mass ejection, .
The star-formation rate is directly proportional to the gas-accretion
rate. At intermediate mass ($M_h \simeq 10^{11} \Msun$), there is not
enough hot gas in equilibrium in the dark-matter potential well (owing
to the strong feedback), and therefore the cooling is not efficient.
Consequently, these intermediate-mass galaxies are deficient in fresh
gas, and the star-formation rate becomes lower than observed at
$1< z <3$ (Figs.~\ref{sfr_function} and \ref{Madau_plot}). In
\cite{Guo_2011} and \cite{Henriques_2013}, the lack of gas in the hot phase is
compensated for by the reincorporation of the gas ejected previously that
was stored in a passive reservoir.\\

In the delayed star formation model $m_4$, the
amount of the star formation rate occurring in merger events and
amplified by the boost factor is higher than in $m_1$ (see
Fig.~\ref{sfr_function}). Indeed, mergers generate a strong
increase in the star formation activity because of the large amount of
the accreted gas that is in the disc in the no-star-forming
phase and to the merger-induced no-star-forming to
star-forming gas conversion (Eqs. \ref{no-sfg2sfg} and
\ref{boost_factor}). The rapid transformation of gas into stars
allows reaching very high star-formation rates (see
Fig.~\ref{sfr_function}).  At lower redshift,  as in standard
models, the larger amount of stellar mass is formed in the
quasi-steady state mode. Even if, at these epochs, the contribution
of merger events strongly decreases, the highest star formation rates
are still found in post-merger structures.

\section{Discussion and conclusion}
\label{conclusion}

We have presented four galaxy formation models and compared them. We showed that classical models $m_1$ (reference),
$m_2$, and $m_3$ fail to reproduce the faint end of the stellar-mass
function. They over-predict the stellar mass in the low-mass dark
matter haloes ($M_h < 10^{10}~\Msun$). Even when a strong
photoionization and SN-feedback are used (as in $m_2$ and $m_3$),
the models form too many stars in the low-mass range. Moreover,
recent observations indicate that the loading factors
($\dot{M}_{ej}/\dot{M}_{\star}$) are much smaller that those
predicted by such models. A strong SN-feedback generates a strong
decrease in the amount of gas, which has to be compensated for at low $z$,
for example, by reincorporating some gas \citep{Henriques_2013}. Such a
problem in the low-mass structures is invariably present, even in the
most recent SAMs \citep{Guo_2011, Bower_2012, Weinmann_2012} and, as
explained by \cite{Henriques_2013}, can thus be viewed as a generic
problem.\\

On the basis of a comparison between the models for which the stellar-mass functions
and the relationship between $M_h$ and $M_{\star}$ are
reproduced quite well at $z=0$ \citep[e.g.][]{Guo_2011}, it seems that the
problem occurs at high redshift. But, if the relations at
high $z$ are not reproduced, then the history of the structures
populating the relations at lower redshift is not consistent, even if the
stellar-mass functions at these low redshifts seem agrees
with observations.\\

If we consider that the $\Lambda-CDM$ paradigm produces the
correct number of low-mass dark matter haloes, then the star formation
activity has to be strongly reduced in this range of masses to
reproduce the observations. We applied this condition in an
ad-hoc model ($m_4$). This model is based on a two-phase
gaseous disc with, on the one hand the star-forming gas, and
on the other, the no-star-forming gas. We showed that
$m_4$ is in good agreement with a large set of observations, even if
there is a tension with the $z>4$ SFR computed from UV measurements
\citep{Bouwens_2007}.\\ 

A galaxy formation and evolution model that uses a strong
modification of the star formation activity (quenched or limited) has
already been studied by \cite{Bouche_2010}. In their model, the fresh
gas accretion is halted for haloes with masses lower than $M_h =
10^{11}~\Msun$. In our model, such small structures host a very large
amount of no-star-forming gas. That this gas has
been accreted but cannot form star leads to the same result as the
non-accretion model proposed by \cite{Bouche_2010}. The results,
produced by the toy model described in \cite{Bouche_2010}, are in
good agreement with observations (sSFR, their Fig. 4, and the
Tully-Fisher relation, their Fig. 5). This independent study
reinforces the hypothesis of a strong decrease in the star formation
efficiency in low-mass structures at high redshift.

In classical SAMs, the cold gas that can form star is
modelled as a homogeneous component, generally following a decreasing
radial exponential profile. In addition, these models consider that a given
fraction of the gas is available to form stars at any moment. The SFR
applied to this cold gas reservoir is computed following \cite{Kennicutt_1998} empirical law, without knowing the exact
fraction of the gas that can actually be impacted by the process. The
ad hoc model presented in this paper is based on a
no-star-forming gas component that can be seen as the
gaseous fraction that is not in optimal conditions forming stars,
i.e., not above the critical density threshold.\\

Observations show that only a small fraction of the total gas mass ($\simeq 15\%$) is
above the optimal column density threshold, and only a small fraction
($\simeq 15\%$) of this dense gas is in prestellar cores
\citep{Andre_2013a, Andre_2013b}. Indeed, before being in the form
of a prestellar core, the gas in the ISM must follow a continuous
structuration process, from the low density accreted gas to the
highest density regions. This structuration cascade needs time and
obviously, at a given time, all the gas in the disc cannot be
available to form stars. The no-star-forming to
star-forming conversion process may be linked to the global
dynamic of the disc. Indeed in highly-disturbed discs (with
a $V/\sigma_v \in [1 - 10]$), as observed at high $z$
\citep[e.g.][]{Genzel_2006, Genzel_2008, Stark_2008, Cresci_2009} and
seen in hydrodynamic simulations \citep[e.g.][]{Keres_2005,
Dekel_2009a, Dekel_2009b, Khochfar_2009}, star formation occurs only
in a few high-density regions (clumps).\\

In addition to these large scale disturbed dynamics at high $z$, the density
structuration process can be limited by other mechanisms that are
seen at low $z$. The turbulence heating can be one of them.  Even if
the main driver of this turbulence is not clearly understood (shocks,
SN kinetic energy injection, tidal interactions, galaxy collisions),
recent infrared spectroscopic observations at intermediate redshift
($1<z<2$) show that the molecular gas can be dynamically heated by
turbulence \citep[e.g.][]{Guillard_2009, Guillard_2012b, Ogle_2010, Appleton_2013} and is thus not available for star formation.
Hydrodynamic simulations by \cite{Bournaud_2010} also show that the
SN energy injection disrupts the dense regions on the smallest scales
(ten to hundred parsecs) from the typical size of star-forming
filaments in the ISM to a typical disc-scale height at $z \simeq 2$.

The main idea behind the no-star-forming gas
reservoir is that, at any given time, only a fraction of the gas can
form stars. Indeed even if Kennicutt's law gives a relation
between the gas content (mass), the geometry (galaxy size), and the
star formation activity, it does not give any information about the
gas fraction that is turned into stars. In a galaxy, a set of
highly concentrated star-forming regions that affect a small
amount of the total gas mass can be compatible with the Kennicutt's
law (sum of SFR in all regions). However, our ad-hoc model $m_4$
leads to a fraction of no-star-forming gas at $z=0$ that is
greater than given by observations.
Our analysis can also be compared to
previous works that introduced a significant fraction of gas that
cannot form stars
\citep[e.g.][]{Pfenniger_Combes_1994a,Pfenniger_Combes_1994b}. In
this context, the study of the gas dynamics and states (with ALMA for
instance) will be a key point in understanding the regulation
of star formation in galaxies.

To model the two-phase disc, we assumed that the accreted gas
is composed of a large fraction (99\%) of no-star-forming
gas, that is progressively converted in the star-forming
phase. The conversion rate (Eq. \ref{no-sfg2sfg}) has not been
defined to explicitly and physically follow the structuration process
but is calibrated to reproduce the stellar-to-halo mass-relation
(SHMR) \citep[e.g.][]{Leauthaud_2012, Moster_2010, Behroozi_2010,
Bethermin_2012a}. This formulation has no other purpose than to
highlight the order of magnitude of the regulation process that has
to be introduced. To reproduce Kennicutt's law on a galaxy scale,
even when a large fraction of the gas can not form stars, we have artificially concentrated the
star-forming gas in the centre of the disc where the density is
the highest. This ad-hoc modification leads to very good
results mainly for the stellar-mass functions and sSFR. On the other
hand, the SFR distribution predicted by this model strongly disagrees with SFR measurements at $z>4$. It shows
the need for an explicit description of the density structuration
process. The gas-mass function predicted by the ad-hoc
model may indicate that galaxies have a gas content that is too
large, even if the comparison with observations is difficult because the
total gas mass function is not known. In the future measuring the gas mass function will be a key observable that will constrain 
the balance between the ejection process and gas regulation in
galaxies.

In a forthcoming work, we plan to follow the structuration of the
gas, using a semi-analytical approach from the largest scales ($r
>h$, the disc-scale height), where the dynamics are governed by
(quasi-)2D turbulence, to the smallest scales ($r <h$), where the star
formation process occurs in 3D molecular clouds \citep{Romeo_2010,
Bournaud_2010}. In current SAMs, the SFR is only
linked to the gas component and to the overall galaxy dynamics
($t_{dyn}$). This approach is not able to correctly describe the
regulation of the star formation. Galaxy discs are complex structures
based on multi-fluid interactions on multi-scales (2D, 3D;
\cite{Shi_2011}). The connexions between the stellar and gas
components generate instabilities that participate in the ISM
structuration \citep[e.g.][]{Jog_1984, Hoffmann_2012}. Star formation
and disc instabilities are thus strongly linked. A better description
of these ISM structuration mechanisms is essential for a better
understanding of the star formation regulation processes.

\section*{Acknowledgments}
The authors thank Jean-loup Puget and Pierre Guillard for very useful
comments and discussions. We also thank the referee for a very helpful report. We acknowledge financial support from the
"Programme National de Cosmologie and Galaxies" (PNCG) of CNRS/INSU,
France.

\bibliographystyle{aa} 
\bibliography{aa3}

\begin{appendix}

\section{Cosmic variance}
\label{cosmic_variance}

\begin{figure}[!h]
\begin{center}
 \includegraphics[scale =
0.7]{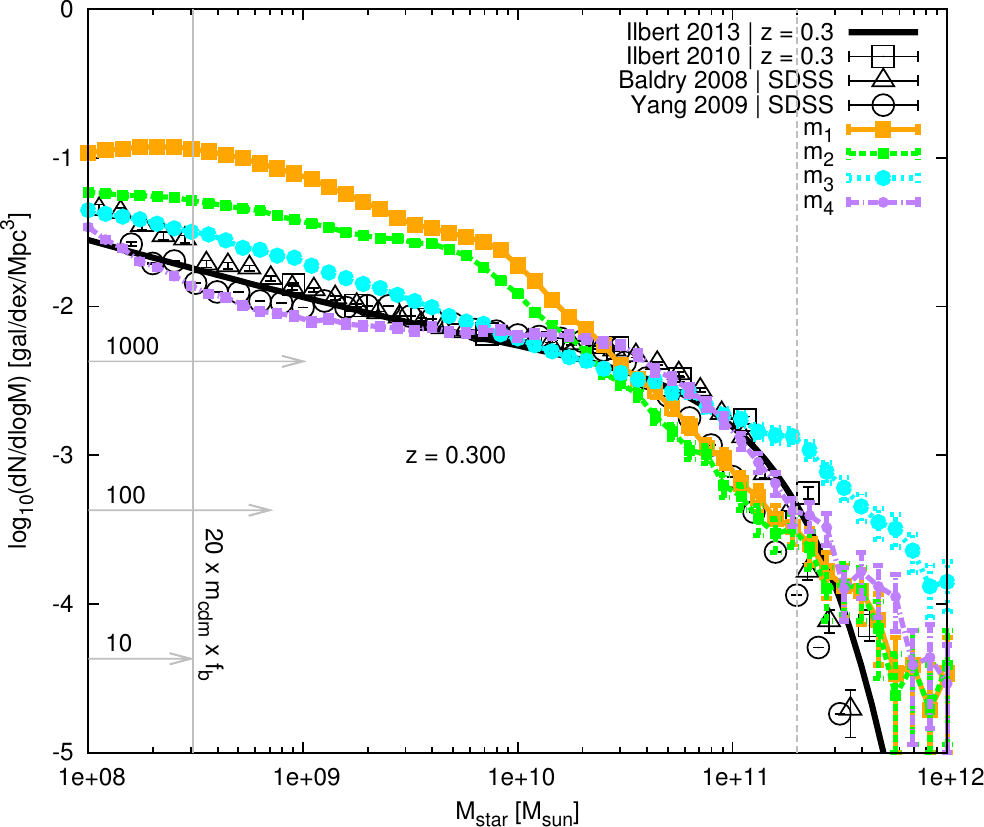}
  \caption{\tiny{Stellar mass function predicted by our different models
at $z=0.3$. The colour code is given in Table 1. We compare our results with
\cite{Ilbert_2010, Ilbert_2013} (squares), \cite{Yang_2009} (circles)
and \cite{Baldry_2008} (triangles) observations. Horizontal arrows
show the link between the density and the number of haloes in our
simulation volume. We show in this figure the error bars related to
cosmic variance.}}
  \label{cosmic_variance_stellar_mass}
\end{center}
\end{figure}

For the sake of clarity, errors bars due to cosmic variance in
the stellar mass function are not plotted in Fig.~\ref{mass_function}. We have indicated a threshold above which
uncertainties become larger than the variations observed for the
different models. For illustration, we show in
Fig.~\ref{cosmic_variance_stellar_mass} the stellar mass functions
predicted by the models at $z=0.3$, where we have added the error bars
linked to the cosmic variance.

\section{The photoionization filtering masses}
\label{filtering _mass}
We used two different models of photoionization based on two
definitions of the filtering mass. We recall here the expressions
that we have used. \\

\subsection{Prescription for $m_1$}
For the \cite{Okamoto_2008} model, we used an expression deduced from a
fit of the filtering mass evolution given in their Fig. 15:
\begin{center}
  \begin{equation}
    M_c(z) =  6\times 10^{9}h_0^{-1}exp\left(-0.7z\right)~\Msun.
  \end{equation}
\end{center}

\subsection{Prescription for $m_2$}

In \cite{Gnedin_2000} and \cite{Kravtsov_2004}, the filtering mass is
defined as a function of the expansion factor ($a$): 
  \begin{equation}
    M_c(a) = 2.5\times
10^{11}h^{-1}\Omega_m^{-1/2}\mu^{-3/2}f(a)~\Msun
    \label{MB_dist}
  \end{equation}
where $f(a)$ is the conditional function given in Eq.
\ref{condi_function}. As explained in \cite{Kravtsov_2004}, the best
agreement with hydrodynamic simulations is obtained with $\alpha=6$.
\onecolumn

\begin{equation}
\tiny
f(a) = \left\{
  \begin{array}{ll}

\dfrac{3a}{(2+\alpha)(5+2\alpha)\left(\dfrac{a}{a_0}\right)^{\alpha}}
& \mbox{: if a $\le a_0$} \\
    & \\

\dfrac{3}{a}\left[a_0^2\left[\dfrac{1}{2+\alpha}-\dfrac{2(a/a_0)^{-1/2}}{5+2\alpha}\right]+\dfrac{a^2}{10}-\dfrac{a_0^2}{10}\left[5-4(a/a_0)^{-1/2}\right]\right]
& \mbox{ : if $a_0 \le a \le a_r$}\\
& \\

\dfrac{3}{a}\left[a_0^2\left[\dfrac{1}{2+\alpha}-\dfrac{2(a/a_0)^{-1/2}}{5+2\alpha}\right]+\dfrac{a_r^2}{10}\left[5-4(a/a_r)^{-1/2}\right]-\dfrac{a_0^2}{10}\left[5-4(a/a_0)^{-1/2}\right]+\dfrac{aa_r}{3}-\dfrac{a_r^2}{3}\left[3-2(a/a_r)^{-1/2}\right]\right]
& \mbox{ : if $\ge a_r$}
  \end{array}\right.
\label{condi_function}
\end{equation}

\section{Data repository}
Outputs from all models are available CDS platform \verb?\http://cdsweb.u-strasbg.fr/cgi-bin/qcat?J/A+A/?. Data are distributed under \verb?*.fits? format and are therefore compatible with the \verb?TOPCAT? software (\verb?http://www.star.bris.ac.uk/~mbt/topcat/?). 

\end{appendix}

\end{document}